\documentstyle[prd,preprint,tighten,aps,eqsecnum,floats,%
amssymb,amsfonts,newlfont,graphicx]{revtex}
\begin{document}
\draft
\preprint{
\begin{tabular}{r}
KIAS-P99074
\\
BROWN-HET-1200
\\
DFTT 43/99
\\
hep-ph/9908513
\\
\end{tabular}
}
\title{Matter Effects in Four-Neutrino Mixing}
\author{David Dooling}
\address{Department of Physics,
Brown University, Providence RI 02912, USA, and
\\
School of Physics, Korea Institute for Advanced Study,
Seoul 130-012, Korea}
\author{Carlo Giunti}
\address{INFN, Sez. di Torino, and Dip. di Fisica Teorica,
Univ. di Torino, I--10125 Torino, Italy, and
\\
School of Physics, Korea Institute for Advanced Study,
Seoul 130-012, Korea}
\author{Kyungsik Kang}
\address{Department of Physics,
Brown University, Providence RI 02912, USA, and
\\
School of Physics, Korea Institute for Advanced Study,
Seoul 130-012, Korea}
\author{Chung W. Kim}
\address{School of Physics, Korea Institute for Advanced Study,
Seoul 130-012, Korea, and
\\
Dept. of Physics $\&$ Astronomy,
The Johns Hopkins University,
Baltimore, MD 21218, USA}
\date{\today}
\maketitle
\begin{abstract}
We present a formalism for the description
of oscillations
in matter of solar, atmospheric and long-baseline neutrinos
in the two schemes of four-neutrino mixing
that are allowed by the results of all existing
 neutrino oscillation experiments.
\end{abstract}

\section{Introduction}
\label{Introduction}

The evidence in favor of neutrino oscillations found
in atmospheric
\cite{SK-atm,exp-atm}
and
solar
\cite{exp-sun}
neutrino experiments
and in the LSND accelerator neutrino experiment \cite{LSND}
implies the existence of at least three independent neutrino
mass-squared differences
(see \cite{Giunti-baksan-99} for a simple proof):
\begin{eqnarray}
&
\Delta{m}^2_{\mathrm{sun}} \sim 10^{-10} \, \mathrm{eV}^2
\;
\mbox{(VO)}
\qquad
\mbox{or}
\qquad
\Delta{m}^2_{\mathrm{sun}} \sim 10^{-6} - 10^{-4} \, \mathrm{eV}^2
\;
\mbox{(MSW)}
\,,
&
\label{dm2-sun}
\\
&
\Delta{m}^2_{\mathrm{atm}} \sim 10^{-3} - 10^{-2} \, \mathrm{eV}^2
\,,
&
\label{dm2-atm}
\\
&
\Delta{m}^2_{\mathrm{LSND}} \sim 1 \, \mathrm{eV}^2
\,.
&
\label{dm2-LSND}
\end{eqnarray}
The two possibilities
(see \cite{analysis-sun})
for the solar
mass-squared difference
$\Delta{m}^2_{\mathrm{sun}}$
correspond
to the vacuum oscillation (VO) solution and
to the MSW effect \cite{MSW}, respectively.

The three independent $\Delta{m}^2$'s in Eqs. (\ref{dm2-sun})--(\ref{dm2-LSND})
imply the existence of at least four massive neutrinos.
In this paper we consider the minimal possibility of
four massive neutrinos
\cite{four-models,four-phenomenology,%
BGKP-96,BGG-AB-96,Okada-Yasuda-97,%
BGG-bounds-98,BGG-CP-98,BGGS-BBN-98,BGGS-AB-99}.
In this case,
in addition to the three active neutrinos $\nu_e$, $\nu_\mu$, $\nu_\tau$
in the flavor basis,
there is a sterile neutrino $\nu_s$
that does not take part in
standard weak interactions (see \cite{sterile}). 

The results of solar and atmospheric neutrino experiments
have been analyzed by several groups
(see
\cite{analysis-sun,SK-atm,Gonzalez-Garcia-atm-analysis-99,Fogli-Lisi-atm}
and references therein)
under the assumption
of transitions into active neutrinos
($\nu_e\to\nu_{\mu,\tau}$ for solar neutrinos
and
$\nu_\mu\to\nu_\tau$
or
$\nu_\mu\rightleftarrows\nu_e$ for atmospheric neutrinos)
or into sterile neutrinos
($\nu_e\to\nu_s$ for solar neutrinos
and
$\nu_\mu\to\nu_s$ for atmospheric neutrinos).
However,
in the case of four-neutrino mixing
transitions into active and sterile neutrinos
can take place simultaneously for both solar and atmospheric neutrinos.
In this paper we present a formalism for the calculation of
the transition probabilities into active and sterile neutrinos
of solar $\nu_e$'s and atmospheric $\nu_\mu$'s,
taking into account the matter effects in the Sun
and in the Earth
(see \cite{Mikheev-Smirnov-uspekhi-87,Bilenky-Petcov-RMP-87,%
Kuo-Pantaleone-RMP-89,CWKim-book,BGG-review-98}).
The derivation of this formalism follows a method
similar to the one used to develop the formalism that describes
neutrino oscillations in the framework of three-neutrino mixing
\cite{Kuo-Pantaleone-PRL-86,Kuo-Pantaleone-RMP-89,%
Shi-Schramm-92,Fogli-Lisi-sun,Narayan-three-nu,GKM-atm-98,%
Akhmedov-Dighe-Lipari-Smirnov-parametric-99}.

It has been shown
\cite{BGG-AB-96,Barger-variations-98,BGGS-AB-99}
that there are only
two schemes with four-neutrino mixing
that can accommodate the results of all neutrino oscillation experiments:
\begin{equation}
\mbox{(A)}
\qquad
\underbrace{
\overbrace{m_1 < m_2}^{\mathrm{atm}}
\ll
\overbrace{m_3 < m_4}^{\mathrm{sun}}
}_{\mathrm{LSND}}
\,,
\qquad
\mbox{(B)}
\qquad
\underbrace{
\overbrace{m_1 < m_2}^{\mathrm{sun}}
\ll
\overbrace{m_3 < m_4}^{\mathrm{atm}}
}_{\mathrm{LSND}}
\,.
\label{AB}
\end{equation}
These two spectra are characterized by the presence of two pairs
of close masses separated by a gap of about 1 eV
which provides the mass-squared difference
$ \Delta{m}^2_{\mathrm{LSND}} = m_4^2 - m_1^2 $
responsible for the oscillations observed in the LSND experiment.
In the scheme A,
we have
$ \Delta{m}^2_{\mathrm{atm}} = m_2^2 - m_1^2 $
and
$ \Delta{m}^2_{\mathrm{sun}} = m_4^2 - m_3^2 $,
whereas in scheme B,
$ \Delta{m}^2_{\mathrm{atm}} = m_4^2 - m_3^2 $
and
$ \Delta{m}^2_{\mathrm{sun}} = m_2^2 - m_1^2 $.
The phenomenology of neutrino oscillations in these two schemes is identical.
Hence,
for simplicity we shall consider in the following only the scheme B,
but all the results are also valid in the scheme A
and can be obtained by the exchange $ 1,2 \leftrightarrows 3,4 $
of the mass eigenstate indices.
The mass spectrum is characterized by the hierarchy
\begin{equation}
\Delta{m}^2_{\mathrm{sun}} = \Delta_{21}
\ll
\Delta{m}^2_{\mathrm{atm}} = \Delta_{43}
\ll
\Delta{m}^2_{\mathrm{LSND}} = \Delta_{41}
\,,
\label{dm2}
\end{equation}
with
$ \Delta_{kj} \equiv m_k^2 - m_j^2 $.

It has been shown in \cite{Okada-Yasuda-97,BGGS-BBN-98}
that the upper bound $ N_\nu^{\mathrm{BBN}} < 4 $
for the effective number of neutrinos
in Big--Bang Nucleosynthesis
implies that
$|U_{s3}|^2+|U_{s4}|^2$
in the scheme B
is very small.
In this case, atmospheric neutrinos oscillate only in the
$\nu_\mu\to\nu_\tau$ channel
and solar neutrino oscillate only in the
$\nu_e\to\nu_s$ channel.
However,
the value of the upper bound for $N_\nu^{\mathrm{BBN}}$
is rather controversial
\cite{BBN-Nnu}
and some authors argue that
$N_\nu^{\mathrm{BBN}}$
could be larger than four.
Moreover,
a large value of
$|U_{s3}|^2+|U_{s4}|^2$
may be compatible with
the upper bound $ N_\nu^{\mathrm{BBN}} < 4 $
if a large lepton asymmetry is generated by
$\nu_\tau\leftrightarrows\nu_s$
oscillations at times
much earlier than the time of Big--Bang Nucleosynthesis
\cite{Foot-Volkas-BBN,Shi-Fuller-BBN,DiBari-Lipari-Lusignoli-BBN-99}.
In this paper we do not put any constraint on the
possible value of
$|U_{s3}|^2+|U_{s4}|^2$.
A global analysis of all neutrino oscillation data
with the formalism presented here could
lead to the determination of the values of
$|U_{s3}|$ and $|U_{s4}|$,
which 
regulate the neutrino oscillations occurring in the early universe.

The plan of the paper is as follows:
in Section \ref{Matter effects}
we introduce the general formulas for the description of
neutrino oscillations in matter
in four-neutrino schemes;
in Section \ref{Solar neutrinos}
we derive the survival probability of solar $\nu_e$'s
and the probabilities of
$\nu_e\to\nu_s$ and $\nu_e\to\nu_{\mu,\tau}$
transitions
in the Sun;
in Section \ref{Regeneration}
we discuss the
regeneration of solar $\nu_e$'s in the Earth;
in Section \ref{Atmospheric neutrinos}
we present the formalism for the description of the oscillations of
high-energy atmospheric neutrinos in the Earth;
in Section \ref{Long-baseline experiments}
we derive the oscillation probabilities of neutrinos
in long-baseline experiments;
in Section \ref{Conclusions}
we draw our conclusions.

\section{Matter effects}
\label{Matter effects}

In a four-neutrino mixing scheme,
the flavor neutrino fields
$\nu_{\alpha L}$
($\alpha=e,s,\mu,\tau$)
are related
to the fields $\nu_{kL}$ of neutrinos with masses $m_k$
by the relation
\begin{equation}
\nu_{\alpha L} = \sum_{k=1}^4 U_{\alpha k} \, \nu_{kL}
\qquad
(\alpha=e,s,\mu,\tau)
\,,
\label{mixing}
\end{equation}
where $U$ is a $4{\times}4$ unitary mixing matrix.
Neglecting possible CP phases,
any $4\times4$ mixing matrix can be written as a product of
six rotations,
\begin{equation}
U_{12} \,, \quad
U_{13} \,, \quad
U_{14} \,, \quad
U_{23} \,, \quad
U_{24} \,, \quad
U_{34} \,,
\label{Uij}
\end{equation}
where
\begin{equation}
(U_{ij})_{ab}
=
\delta_{ab}
+
\left( \cos{\vartheta_{ij}} - 1 \right)
\left( \delta_{ia} \delta_{ib} + \delta_{ja} \delta_{jb} \right)
+
\sin{\vartheta_{ij}}
\left( \delta_{ia} \delta_{jb} - \delta_{ja} \delta_{ib} \right)
\,.
\label{rotation}
\end{equation}
The order of the product of the matrices (\ref{Uij})
corresponds to a specific parameterization
of the mixing matrix $U$.
We will see that in order to study matter effects
in the oscillations of the solar and atmospheric neutrinos,
it is convenient to use different
parameterizations of the mixing matrix $U$.

The evolution of the amplitudes
$\psi_\alpha$ ($\alpha=e,s,\mu,\tau$)
of the
neutrino flavor eigenstates
$|\nu_\alpha\rangle$
in matter is governed by the
MSW equation
\cite{MSW}
\begin{equation}
i \, \frac{ \mathrm{d}}{ \mathrm{d} x } \, \Psi
=
\mathcal{H} \, \Psi
\,,
\label{evolution1}
\end{equation}
where
\begin{equation}
\Psi
=
\left(
\psi_e,
\psi_s,
\psi_\mu,
\psi_\tau
\right)^T
\,
\label{flavor-basis}
\end{equation}
(for example,
apart from a phase,
$|\nu_e\rangle\mapsto(1,0,0,0)^T$,
etc.)
and
$\mathcal{H}$ is the effective Hamiltonian
\begin{equation}
\mathcal{H}
=
\frac{1}{2p}
\left( \mathcal{M}^2 + \mathcal{A} \right)
\,.
\label{hamiltonian1}
\end{equation}
Here $p$ is the neutrino momentum,
$\mathcal{M}^2$
is the mass-squared matrix in the flavor basis,
\begin{equation}
\mathcal{M}^2
=
U \, \mathcal{M}_0^2 \, U^\dagger
\,,
\label{mass1}
\end{equation}
with
$ \mathcal{M}_0^2 = \mathrm{diag}(m_1^2,m_2^2,m_3^2,m_4^2) $,
and $\mathcal{A}$ is the matrix
\begin{equation}
\mathcal{A} = \mathrm{diag}(A_{CC},-A_{NC},0,0)
\,,
\label{A}
\end{equation}
with
$ A_{CC} = 2 p V_{CC} $
and
$ A_{NC} = 2 p V_{NC} $
(see \cite{Mikheev-Smirnov-uspekhi-87,Bilenky-Petcov-RMP-87,%
Kuo-Pantaleone-RMP-89,CWKim-book,BGG-review-98}).
Here
$V_{CC}$ and $V_{NC}$
are the charged-current and neutral-current matter potentials
 \begin{eqnarray}
V_{CC}
&=&
\sqrt{2} \, G_F \, N_e
=
7.63 \times 10^{-14}
\left(
\frac{ N_e }{ N_A \, \mathrm{cm}^{-3} }
\right)
\mathrm{eV}
\,,
\label{VCC}
\\
V_{NC}
&=&
- \frac{1}{2} \sqrt{2} G_F N_n
\,,
\label{VNC}
\end{eqnarray}
where
$G_F$ is the Fermi constant,
$N_A$ is the Avogadro number
and
$N_e$ and $N_n$ are the electron and neutron number densities, respectively.
These number densities are
given by
$N_e = \rho N_A Y_e$
and
$N_n = \rho N_A (1-Y_e)$,
where
$\rho$ is the matter density in g/cm$^3$
and $Y_e$ is the electron number fraction.
In the case of antineutrinos
$A_{CC}$ and $A_{NC}$
must be replaced by
$\overline{A}_{CC}=-A_{CC}$ and $\overline{A}_{NC}=-A_{NC}$,
respectively.
In Eq. (\ref{evolution1}) we have subtracted a common phase
generated by $ p + V_{NC} $
that has no effect on flavor transitions.

The flavor basis (\ref{flavor-basis}) is related to the mass basis
\begin{equation}
\Psi^0
=
\left(
\psi_1,
\psi_2,
\psi_3,
\psi_4
\right)^T
\label{mass-basis}
\end{equation}
by
\begin{equation}
\Psi
=
U \, \Psi^0
\,.
\label{mixing-of-amplitudes}
\end{equation}
Here
$\psi_k$ ($k=1,2,3,4$)
are the amplitudes of the mass eigenstate neutrinos
$|\nu_k\rangle$,
which are related to the flavor eigenstates by
\begin{equation}
|\nu_\alpha\rangle
=
\sum_{k=1}^4
U_{{\alpha}k} \, |\nu_k\rangle
\qquad
(\alpha=e,s,\mu,\tau)
\,.
\label{mixing-of-states}
\end{equation}

\section{Solar neutrinos}
\label{Solar neutrinos}

Because the solar neutrino oscillations are due to the
mass-squared difference $\Delta_{21}$,
a convenient parameterization
of the mixing matrix $U$
for studying the matter effects is
\begin{equation}
U = U_{34} \, U_{24} \, U_{23} \, U_{14} \, U_{13} \, U_{12}
\,.
\label{U-sun}
\end{equation}
Furthermore,
it is convenient to work in the rotated basis
\begin{equation}
\Psi'
=
{U'}^\dagger \, \Psi
\,,
\label{rotated-basis-1}
\end{equation}
with
\begin{equation}
U'
=
U_{34} \, U_{24} \, U_{23} \, U_{14} \, U_{13}
\label{Up}
\end{equation}
and
\begin{equation}
\Psi'
=
\left(
\psi'_1,
\psi'_2,
\psi'_3,
\psi'_4
\right)^T
\,.
\label{rotated-basis}
\end{equation}
Here
$\psi'_k$ ($k=1,2,3,4$)
are the amplitudes of the rotated neutrino states
$
|\nu'_k\rangle
=
\sum_\alpha U'_{{\alpha}k} |\nu_\alpha\rangle
$.
The relation between the rotated basis (\ref{rotated-basis-1})
and the mass basis (\ref{mass-basis}) is
\begin{equation}
\Psi'
=
U_{12} \, \Psi^0
\,,
\label{rotated-basis-2}
\end{equation}
from which it is clear that
$\psi'_3=\psi_3$
and
$\psi'_4=\psi_4$.

The evolution equation for $\Psi'$ is
\begin{equation}
i \, \frac{ \mathrm{d} }{ \mathrm{d} x } \, \Psi'
=
\mathcal{H}' \, \Psi'
\,,
\label{evolution2}
\end{equation}
with
\begin{equation}
\mathcal{H}'
=
\frac{1}{2p}
\left( U_{12} \, \mathcal{M}_0^2 \, U_{12}^\dagger + {U'}^\dagger \, \mathcal{A} \, U' \right)
\,.
\label{hamiltonian2}
\end{equation}
Because the rotation $U_{34}$ does not have any effect on the matrix $\mathcal{A}$,
the effective Hamiltonian
$\mathcal{H}'$ is independent of the mixing angle $\vartheta_{34}$.
On the other hand,
since the matrix $\mathcal{A}$ is not invariant under
the rotations
$U_{13}$, $U_{14}$, $U_{23}$, $U_{24}$,
the general form of $\mathcal{H}'$
is very complicated.
In order to seek a simpler form,
we use the fact that
the negative results of short-baseline $\bar\nu_e$
disappearance experiments
lead to small mixing angles
$\vartheta_{13}$
and
$\vartheta_{14}$
\cite{BGKP-96,BGG-AB-96}.
Indeed,
for $\Delta_{41}$ in the LSND allowed region,
the results of the Bugey $\bar\nu_e$ disappearance experiment
\cite{Bugey-95}
imply that
\begin{equation}
\sin^2{\vartheta_{13}} + \sin^2{\vartheta_{14}}
=
|U_{e3}|^2 + |U_{e4}|^2 \lesssim 10^{-2}
\,.
\label{small}
\end{equation}
Therefore,
in the following
we neglect the two small angles
$\vartheta_{13}$
and
$\vartheta_{14}$
whose contribution to the solar neutrino transitions is negligible.
Let us, however, emphasize that
although
$\vartheta_{13}$
and
$\vartheta_{14}$
are both small,
at least one of the two must be different from zero
in order to generate the
$\bar\nu_\mu\to\bar\nu_e$
and
$\nu_\mu\to\nu_e$
oscillations
observed in the LSND experiment,
whose amplitude is
$
A_{e\mu}
=
4
\left| \sum_{k=3,4} U_{ek}^* U_{{\mu}k} \right|^2
$
\cite{BGKP-96,BGG-AB-96}.

With the approximation
$\vartheta_{13}=\vartheta_{14}=0$,
the mixing matrix reduces to
$ U = U_{34} \, U_{24} \, U_{23} \, U_{12} $
and its explicit form is
\begin{equation}
U
=
\left(
\begin{array}{cccc} \scriptstyle
c_{\vartheta_{12}}
& \scriptstyle
s_{\vartheta_{12}}
& \scriptstyle
0
& \scriptstyle
0
\\ \scriptstyle
- s_{\vartheta_{12}} c_{\vartheta_{23}} c_{\vartheta_{24}}
& \scriptstyle
c_{\vartheta_{12}} c_{\vartheta_{23}} c_{\vartheta_{24}}
& \scriptstyle
s_{\vartheta_{23}} c_{\vartheta_{24}}
& \scriptstyle
s_{\vartheta_{24}}
\\ \scriptstyle
s_{\vartheta_{12}}
( c_{\vartheta_{23}} s_{\vartheta_{24}} s_{\vartheta_{34}}
+ s_{\vartheta_{23}} c_{\vartheta_{34}} )
& \scriptstyle
- c_{\vartheta_{12}}
( s_{\vartheta_{23}} c_{\vartheta_{34}}
+ c_{\vartheta_{23}} s_{\vartheta_{24}} s_{\vartheta_{34}} )
& \scriptstyle
c_{\vartheta_{23}} c_{\vartheta_{34}}
- s_{\vartheta_{23}} s_{\vartheta_{24}} s_{\vartheta_{34}}
& \scriptstyle
c_{\vartheta_{24}} s_{\vartheta_{34}}
\\ \scriptstyle
s_{\vartheta_{12}}
( c_{\vartheta_{23}} s_{\vartheta_{24}} c_{\vartheta_{34}}
- s_{\vartheta_{23}} s_{\vartheta_{34}} )
& \scriptstyle
c_{\vartheta_{12}}
( s_{\vartheta_{23}} s_{\vartheta_{34}}
- c_{\vartheta_{23}} s_{\vartheta_{24}} c_{\vartheta_{34}} )
& \scriptstyle
-
( c_{\vartheta_{23}} s_{\vartheta_{34}}
+ s_{\vartheta_{23}} s_{\vartheta_{24}} c_{\vartheta_{34}} )
& \scriptstyle
c_{\vartheta_{24}} c_{\vartheta_{34}}
\end{array}
\right)
\,,
\label{U-sun-explicit}
\end{equation}
with
$ c_\vartheta \equiv \cos\vartheta $
and
$ s_\vartheta \equiv \sin\vartheta $.
One can easily see that the mixing matrix (\ref{U-sun-explicit})
entails the following limiting cases:
\begin{description}
\item[$\vartheta_{23}=\vartheta_{24}=0.$]
The mixing matrix $U$ assumes a block-diagonal form
in which
$\nu_e$, $\nu_s$
are linear combinations of
$\nu_1$, $\nu_2$
alone
and
$\nu_\mu$, $\nu_\tau$
are linear combinations of
$\nu_3$, $\nu_4$
alone.
Therefore, this case corresponds to pure
$\nu_e\to\nu_s$ transitions for the solar neutrinos
and pure
$\nu_\mu\to\nu_\tau$ transitions for the atmospheric neutrinos.
\item[$\vartheta_{23}=\pi/2.$]
In this case,
$U_{s1}=U_{s2}=0$
and the solar neutrinos undergo
$\nu_e\to\nu_\mu$ and/or $\nu_e\to\nu_\tau$ transitions,
whereas
the atmospheric neutrinos undergo
$\nu_\mu\to\nu_s$ and/or $\nu_\mu\to\nu_\tau$ transitions.
\item[$\vartheta_{23}=\pi/2, \, \vartheta_{34}=\pi/2.$]
The mixing matrix $U$ assumes a block-diagonal form
in which
$\nu_e$, $\nu_\tau$
are linear combinations of
$\nu_1$, $\nu_2$
alone
and
$\nu_\mu$, $\nu_s$
are linear combinations of
$\nu_3$, $\nu_4$
alone.
This case corresponds to pure
$\nu_e\to\nu_\tau$ transitions for the solar neutrinos
and pure
$\nu_\mu\to\nu_s$ transitions for the atmospheric neutrinos.
\item[$\vartheta_{24}=\pi/2.$]
This case is excluded because it yields
$U_{s4}=1$.
Hence,
$\nu_4=\nu_s$
is decoupled and does not participate in neutrino oscillations,
excluding the possibility to explain the atmospheric neutrino data.
\end{description}

With the approximation
$\vartheta_{13}=\vartheta_{14}=0$
the matrix $U'$ is given by
\begin{equation}
U'
=
U_{34} \, U_{24} \, U_{23}
\,,
\label{Ups}
\end{equation}
and
the effective Hamiltonian $\mathcal{H}'$ assumes the simplified form
\begin{equation}
\mathcal{H}'
=
\frac{1}{2p}
\left(
U_{12} \, \mathcal{M}_0^2 \, U_{12}^\dagger
+
U_{23}^\dagger \, U_{24}^\dagger \, \mathcal{A} \, U_{24} \, U_{23}
\right)
\,,
\label{hamiltonian3}
\end{equation}
which explicitly reads as
\begin{equation}
\mathcal{H}'
=
\frac{1}{4p}
\left(
\begin{array}{cccc}
\scriptstyle
\Sigma_{21} - \Delta_{21} c_{2\vartheta_{12}} + 2 A_{CC}
&
\scriptstyle
\Delta_{21} s_{2\vartheta_{12}}
&
\scriptstyle
0
&
\scriptstyle
0
\\
\scriptstyle
\Delta_{21} s_{2\vartheta_{12}}
&
\scriptstyle
\Sigma_{21} + \Delta_{21} c_{2\vartheta_{12}}
- 2 c^2_{\vartheta_{23}} c^2_{\vartheta_{24}} A_{NC}
&
\scriptstyle
- s_{2\vartheta_{23}} c^2_{\vartheta_{24}} A_{NC}
&
\scriptstyle
- c_{\vartheta_{23}} s_{2\vartheta_{24}} A_{NC}
\\
\scriptstyle
0
&
\scriptstyle
- s_{2\vartheta_{23}} c^2_{\vartheta_{24}} A_{NC}
&
\scriptstyle
2 m_3^2 - 2 s^2_{\vartheta_{23}} c^2_{\vartheta_{24}} A_{NC}
&
\scriptstyle
- s_{\vartheta_{23}} s_{2\vartheta_{24}} A_{NC}
\\
\scriptstyle
0
&
\scriptstyle
- c_{\vartheta_{23}} s_{2\vartheta_{24}} A_{NC}
&
\scriptstyle
- s_{\vartheta_{23}} s_{2\vartheta_{24}} A_{NC}
&
\scriptstyle
2 m_4^2 - 2 s^2_{\vartheta_{24}} A_{NC}
\end{array}
\right)
\,,
\label{hamiltonian4}
\end{equation}
with
$ \Sigma_{21} \equiv m_2^2 + m_1^2 $.
This is our desired simpler form of $\mathcal{H}'$.

From the hierarchy (\ref{dm2})
and the values of the first two diagonal elements in $\mathcal{H}'$,
we find that there is a MSW resonance in the Sun for
\begin{equation}
A
=
\Delta_{21} \, \cos{2\vartheta_{12}}
\,,
\label{resonance}
\end{equation}
with
\begin{equation}
A
\equiv
A_{CC} + \cos^2{\vartheta_{23}} \, \cos^2{\vartheta_{24}} \, A_{NC}
\,.
\label{Atilde}
\end{equation}
The resonance condition (\ref{resonance})
has the same form as the well-known resonance condition
for two-generation mixing,
with the replacement of the usual quantity
$A_{CC}$ in the case of $\nu_e\to\nu_{\mu,\tau}$ transitions,
or $A_{CC}+A_{NC}$ in the case of $\nu_e\to\nu_s$ transitions,
by $A_{CC} + \cos^2{\vartheta_{23}} \cos^2{\vartheta_{24}} A_{NC}$.
Indeed,
as explained after Eq. (\ref{U-sun-explicit}),
the two extreme cases of
$\nu_e\to\nu_{\mu,\tau}$
and
$\nu_e\to\nu_s$ transitions
for the solar neutrinos
are recovered in the four-neutrino mixing formalism
for
$\vartheta_{23}=\pi/2$
and
$\vartheta_{23}=\vartheta_{24}=0$,
respectively.

\begin{figure}[t]
\begin{center}
\includegraphics[bb=81 554 447 768,width=0.80\linewidth]{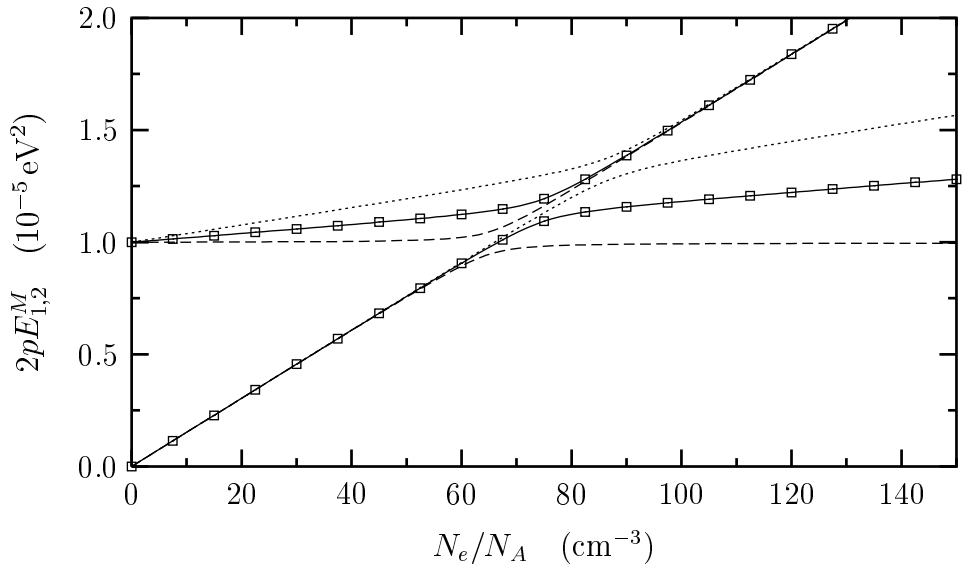}
\caption{ \label{eigen-sun}
The solid lines show the eigenvalues
of the effective Hamiltonian
(\ref{hamiltonian4})
as functions of the electron number density $N_e$
obtained with a numerical diagonalization,
whereas
the open squares show the eigenvalues obtained with Eq. (\ref{eigenvalues}).
We assumed
$ m_1 = 0 $,
$ \Delta_{21} = 1.0 \times 10^{-5} \, \mathrm{eV}^2 $,
$ \Delta_{43} = 1.3 \times 10^{-3} \, \mathrm{eV}^2 $,
$ \Delta_{41} = 1 \, \mathrm{eV}^2 $,
$ \sin^2{2\vartheta_{12}} = 10^{-2} $,
$
\cos^2{\vartheta_{23}}
=
\cos^2{\vartheta_{24}}
=
1/\sqrt{2}
$,
$ p = 1 \, \mathrm{MeV} $
and
$ N_n = N_e/2 $.
The dashed and dotted lines
show the eigenvalues
for
$
\cos{\vartheta_{23}}
=
0
$
and
$
\cos{\vartheta_{23}}
=
\cos{\vartheta_{24}}
=
1
$,
respectively.
}
\end{center}
\end{figure}

Since
$ \Delta_{41} \gg \Delta_{43} \gg \Delta_{21} \sim A $,
the evolution in the Sun of the amplitudes
$\psi'_3$ and $\psi'_4$ is adiabatic and decoupled from the
evolution of the amplitudes
$\psi'_1$ and $\psi'_2$,
which can undergo a MSW resonance.
To lowest order in the power expansion in the small quantities
$
|A_{CC}|/m_3^2
\sim
|A_{CC}|/m_4^2
\sim
|A_{NC}|/m_3^2
\sim
|A_{NC}|/m_4^2
\sim
\Delta_{21}/m_3^2
\sim
\Delta_{21}/m_4^2
$,
the eigenvalues of the effective Hamiltonian $\mathcal{H}'$ are
\begin{equation}
E^M_{2,1}
=
\frac{1}{4p}
\left(
\Sigma_{21}
+ A_{CC} - \cos^2{\vartheta_{23}} \cos^2{\vartheta_{24}} A_{NC}
\pm
\sqrt{
\left( \Delta_{21} \cos{2\vartheta_{12}} - A \right)^2
+
\left( \Delta_{21} \sin{2\vartheta_{12}} \right)^2
}
\right)
\,,
\label{eigenvalues}
\end{equation}
and
$ E^M_3 = E_3 = m_3^2 / 2 p $,
$ E^M_4 = E_4 = m_4^2 / 2 p $.
The accuracy of Eq. (\ref{eigenvalues})
is shown in Fig. \ref{eigen-sun}
in which the solid lines show the eigenvalues
of the effective Hamiltonian
$\mathcal{H}'$
as functions of the electron number density $N_e$
obtained with a numerical diagonalization
and the open squares show
the eigenvalues obtained with Eq. (\ref{eigenvalues}).
The values of the mixing parameters used in drawing Fig. \ref{eigen-sun}
are
$ m_1 = 0 $,
$ \Delta_{21} = 1.0 \times 10^{-5} \, \mathrm{eV}^2 $,
$ \Delta_{43} = 1.3 \times 10^{-3} \, \mathrm{eV}^2 $,
$ \Delta_{41} = 1 \, \mathrm{eV}^2 $,
$ \sin^2{2\vartheta_{12}} = 10^{-2} $,
$
\cos^2{\vartheta_{23}}
=
\cos^2{\vartheta_{24}}
=
1/\sqrt{2}
$,
$ p = 1 \, \mathrm{MeV} $
and we have assumed
$ N_n = N_e/2 $.
The dashed curves in Fig. \ref{eigen-sun}
show the eigenvalues
for
$
\cos{\vartheta_{23}}
=
0
$,
corresponding to pure
$\nu_e\to\nu_{\mu,\tau}$
transitions in the Sun.
On the other hand,
the dotted curves in Fig. \ref{eigen-sun}
show the eigenvalues
for
$
\cos{\vartheta_{23}}
=
\cos{\vartheta_{24}}
=
1
$,
corresponding to pure
$\nu_e\to\nu_s$
transitions in the Sun.
One can clearly see that as
$
\cos^2{\vartheta_{23}}
\cos^2{\vartheta_{24}}
$
increases,
the location of the resonance is shifted towards
higher matter densities.

Apart from the common term $p+V_{NC}$
which has been subtracted from the beginning,
the eigenvalues $E^M_k$ ($k=1,2,3,4$)
are the energies of the energy eigenstates in matter
$|\nu^M_k\rangle$,
whose column matrix of amplitudes
\begin{equation}
\Psi^M
=
\left(
\psi^M_1,
\psi^M_2,
\psi^M_3,
\psi^M_4
\right)^T
\label{states3}
\end{equation}
is given by
\begin{equation}
\Psi^M
=
{U^M_{12}}^\dagger \, \Psi'
\,,
\label{matter-mixing1}
\end{equation}
where
\begin{equation}
(U^M_{12})_{ab}
=
\delta_{ab}
+
\left( \cos{\vartheta_{12}^M} - 1 \right)
\left( \delta_{1a} \delta_{1b} + \delta_{2a} \delta_{2b} \right)
+
\sin{\vartheta_{12}^M}
\left( \delta_{1a} \delta_{2b} - \delta_{2a} \delta_{1b} \right)
\label{U12M}
\end{equation}
is the orthogonal matrix that diagonalizes the 1--2 sector
of the effective Hamiltonian 
$\mathcal{H}'$ in Eq. (\ref{hamiltonian4}).
The value of the effective mixing angle in matter $\vartheta_{12}^M$
is given by
\begin{equation}
\tan 2\vartheta_{12}^M
=
\frac
{ \tan 2\vartheta_{12} }
{ 1 - A / \Delta_{21} \cos{2\vartheta_{12}} }
\,.
\label{theta12m}
\end{equation}
One can clearly see that there is a resonance
when the condition (\ref{resonance})
is satisfied
and the effective mixing angle becomes
$ \vartheta_{12}^M |_{R} = \pi/4 $.
The adiabaticity of the resonance is
characterized by the adiabaticity parameter
(see \cite{BGG-review-98} and references therein)
\begin{equation}
\gamma
=
\left.
\frac
{ E^M_2 - E^M_1 }
{ 2 \left| \mathrm{d} \vartheta_{12}^M / \mathrm{d}x \right| }
\right|_{R}
=
\frac
{ \Delta_{21} \sin^2{2\vartheta_{12}} }
{ 2 p \cos{2\vartheta_{12}}
\left( |\dot{A}| / A \right)_{R} }
\,,
\label{adiabaticity}
\end{equation}
where
$ \dot{A} \equiv \mathrm{d} A / \mathrm{d}x $
and the subscript $R$ denotes values at the resonance point.

The connection between the flavor basis
$\Psi$
and the energy basis in matter $\Psi^M$ is given by
\begin{equation}
\Psi
=
U' \, U^M_{12} \, \Psi^M
=
U^M \, \Psi^M
\,,
\label{matter-mixing2}
\end{equation}
where $ U^M = U' \, U^M_{12} $
is the effective mixing matrix in matter
given by Eq. (\ref{U-sun-explicit})
with the replacement of the vacuum mixing angle $\vartheta_{12}$
with the effective mixing angle in matter $\vartheta_{12}^M$.
Therefore,
the mixing angles $\vartheta_{23}$, $\vartheta_{24}$ and $\vartheta_{34}$
are not modified in matter,
whereas
the mixing angle $\vartheta_{12}$
assumes the effective value $\vartheta_{12}^M$
given by Eq. (\ref{theta12m}).

The energy basis in matter $\Psi^M$ is connected to the mass basis $\Psi^0$
by the relation
\begin{equation}
\Psi^M = {U^M_{12}}^{\dagger} \, U_{12} \, \Psi^0
\,.
\label{energy-mass-basis}
\end{equation}
In particular one can see that
$\psi^M_3=\psi_3$
and
$\psi^M_4=\psi_4$.
Therefore,
the energy eigenstates in matter
$|\nu^M_3\rangle$
and
$|\nu^M_4\rangle$
coincide with the corresponding mass eigenstates
$|\nu_3\rangle$
and
$|\nu_4\rangle$
(energy eigenstates in vacuum)
and their evolution is adiabatic and decoupled from the
evolution of the states
$|\nu^M_1\rangle$
and
$|\nu^M_2\rangle$.

Let us now calculate the probabilities
of flavor transitions for solar neutrinos.
We consider a neutrino produced in the core of the Sun
at the point $x_0$
where,
for appropriate values of $\Delta_{21}$
and of the neutrino momentum $p$,
the matter density is higher than the resonance density given by
Eq. (\ref{resonance}).
Since its state is an electron neutrino state,
we have
\begin{equation}
\Psi(x_0)
=
(1,0,0,0)^T
\,,
\label{initial-amplitudes-flavor}
\end{equation}
and
the initial amplitudes of the energy eigenstates are
\begin{equation}
\psi^M_k(x_0) = U^M_{ek}(x_0)
\qquad
(k=1,2,3,4)
\,,
\label{initial-amplitudes-energy}
\end{equation}
\textit{i.e.}
$ \Psi^M(x_0) = ( \cos{\vartheta^M_{12}}(x_0),
\sin{\vartheta^M_{12}}(x_0), 0 , 0 ) $.
The amplitudes of the mass eigenstates $|\nu_k\rangle$
at the point $x$
in vacuum out of the Sun are
\begin{equation}
\psi_k(x)
=
\sum_{j=1,2}
\phi_k(x-x_R)
\,
T_{kj}
\,
\phi_j(x_R-x_0)
\,
U_{ej}^M(x_0)
\qquad
(k=1,2)
\,,
\label{final-amplitude1}
\end{equation}
and
$\psi_3(x)=\psi_4(x)=0$.
Here $x_R$ is the resonance point,
$\phi_k(x_2-x_1)$
is the phase that the $k^{\mathrm{th}}$
energy eigenstate acquires going from $x_1$ to $x_2$
and $T_{kj}$ is the amplitude of
$\nu_j^M\to\nu_k^M$
transitions in the resonance.
The values of the mass eigenstate indices in Eq. (\ref{final-amplitude1})
are limited to $1,2$
because
$\psi_3(x_0)=\psi_4(x_0)=0$
(due to the fact that $U_{e3}^M=U_{e4}^M=0$),
and
in the resonance
there are no transitions from the states
$|\nu_1^M\rangle$ and $|\nu_2^M\rangle$
to the states
$|\nu_3^M\rangle$ and $|\nu_4^M\rangle$,
that evolve adiabatically because of the decoupling
of the 3--4 sector in the effective Hamiltonian (\ref{hamiltonian4}).

Since the probabilities of
$\nu_j^M\to\nu_k^M$
and
$\nu_k^M\to\nu_j^M$
transitions in the resonance are equal,
we have
\begin{equation}
|T_{12}|^2 = |T_{21}|^2 = P_c
\,,
\qquad
|T_{11}|^2 = |T_{22}|^2 = 1 - P_c
\,,
\label{crossing}
\end{equation}
where $P_c$ is the crossing probability given by
\begin{equation}
P_c
=
\frac{
\exp\left(
- \frac{\pi}{2} \, \gamma \, F
\right)
-
\exp\left(
- \frac{\pi}{2} \, \gamma \, \frac{F}{\sin^2{\vartheta_{12}}}
\right)
}
{
1
-
\exp\left(
- \frac{\pi}{2} \, \gamma \, \frac{F}{\sin^2{\vartheta_{12}}}
\right)
}
\,,
\label{Pc}
\end{equation}
where $\gamma$ is the adiabaticity parameter (\ref{adiabaticity})
and $F$ depends on the slope of $A$ in the resonance
(see \cite{BGG-review-98} and references therein).
For an exponentially varying matter density
$ F = 1 - \tan^2 \vartheta_{12} $.
This expression for $F$ is a good approximation for the Sun.

\begin{figure}[t]
\begin{center}
\includegraphics[bb=40 162 430 801,height=0.84\textheight]{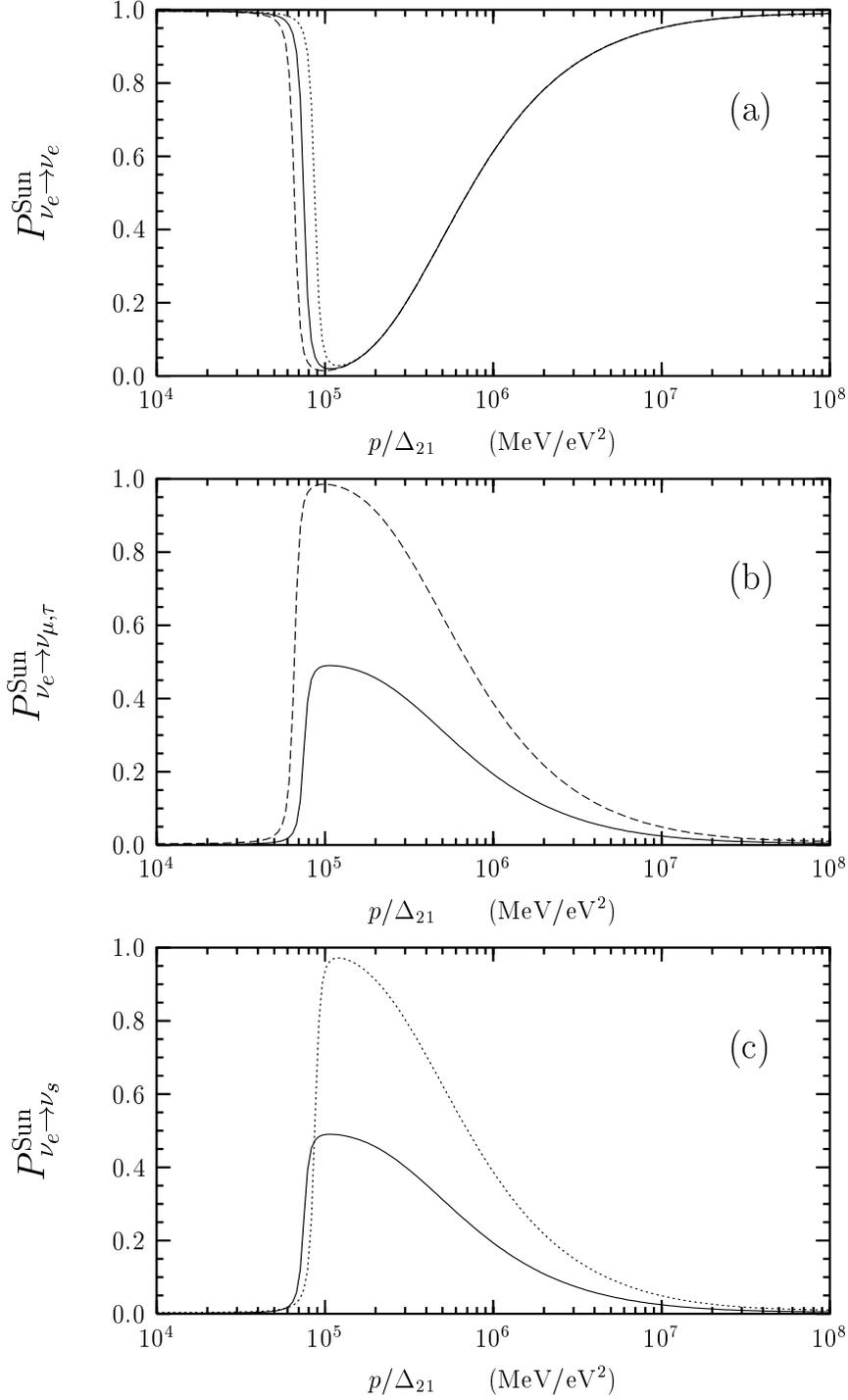}
\caption{ \label{prob-sun}
Survival probability of solar $\nu_e$'s (a)
and probabilities of
$\nu_e\to\nu_{\mu,\tau}$ (b)
and
$\nu_e\to\nu_s$ (c)
transitions in the Sun
for neutrinos produced in the center of the Sun
as functions of $p/\Delta_{21}$,
for
$ \sin^2 2\vartheta_{12} = 6 \times 10^{-3} $.
The dashed, solid and dotted curves
correspond to
$ \cos^2{\vartheta_{23}} \cos^2{\vartheta_{24}} = 0 $
(only $\nu_e\to\nu_{\mu,\tau}$ transitions),
$ \cos^2{\vartheta_{23}} \cos^2{\vartheta_{24}} = 0.5 $
and
$ \cos^2{\vartheta_{23}} \cos^2{\vartheta_{24}} = 1 $
(only $\nu_e\to\nu_s$ transitions),
respectively.}
\end{center}
\end{figure}

From Eq. (\ref{final-amplitude1}),
the amplitude of the flavor state $|\nu_\alpha\rangle$
at the point $x$
in vacuum out of the Sun is
\begin{equation}
\psi_\alpha(x)
=
\sum_{k,j=1,2}
U_{{\alpha}k}
\,
\phi_k(x-x_R)
\,
T_{kj}
\,
\phi_j(x_R-x_0)
\,
U_{ej}^M(x_0)
\,,
\label{final-amplitude2}
\end{equation}
and the probabilities of
$\nu_e\to\nu_\alpha$ transitions
($\alpha=e,s,\mu,\tau$)
are given by
\begin{equation}
P^{\mathrm{Sun}}_{\nu_e\to\nu_\alpha}
=
| \psi_\alpha |^2
=
\left|
\sum_{k,j=1,2}
U_{{\alpha}k}
\,
\phi_k(x-x_R)
\,
T_{kj}
\,
\phi_j(x_R-x_0)
\,
U_{ej}^M(x_0)
\right|^2
\,.
\label{prob1}
\end{equation}
Because the phases oscillate rapidly over the neutrino energy spectrum
and
over the spread of
the source--resonance and resonance--detector distances,
the interference terms in this expression are practically not observable.
The observable averaged probabilities are given by
\begin{eqnarray}
P^{\mathrm{Sun}}_{\nu_e\to\nu_\alpha}
&=&
\sum_{k,j=1,2}
|U_{{\alpha}k}|^2
\,
|T_{kj}|^2
\,
|U_{ej}^M(x_0)|^2
\nonumber
\\
&=&
\left(
|U_{{\alpha}1}|^2 |U_{e1}^M(x_0)|^2
+
|U_{{\alpha}2}|^2 |U_{e2}^M(x_0)|^2
\right)
\left( 1 - P_c \right)
\nonumber
\\
&&
+
\left(
|U_{{\alpha}1}|^2 |U_{e2}^M(x_0)|^2
+
|U_{{\alpha}2}|^2 |U_{e1}^M(x_0)|^2
\right)
P_c
\,.
\label{prob2}
\end{eqnarray}
With this expression and
Eq. (\ref{U-sun-explicit})
for the mixing matrix in vacuum
and the corresponding expression for the effective mixing matrix
at the production point $x_0$
(given by Eq. (\ref{U-sun-explicit})
with the replacement
$\vartheta_{12}\to\vartheta_{12}^M(x_0)$),
it is straightforward to calculate the
$\nu_e$ survival probability and the
probability of $\nu_e\to\nu_s$ transitions:
\begin{eqnarray}
&&
P^{\mathrm{Sun}}_{\nu_e\to\nu_e}
=
\frac{1}{2}
+
\left( \frac{1}{2} - P_c \right) \cos{2\vartheta_{12}} \, \cos{2\vartheta_{12}^M}(x_0)
\,,
\label{Pee}
\\
&&
P^{\mathrm{Sun}}_{\nu_e\to\nu_s}
=
\cos^2{\vartheta_{23}} \cos^2{\vartheta_{24}}
\left[
\frac{1}{2}
-
\left( \frac{1}{2} - P_c \right) \cos{2\vartheta_{12}} \, \cos{2\vartheta_{12}^M}(x_0)
\right]
\,.
\label{Pes}
\end{eqnarray}
One can see that the survival probability has the same form as in the
two-generation case
(see \cite{Mikheev-Smirnov-uspekhi-87,Bilenky-Petcov-RMP-87,%
Kuo-Pantaleone-RMP-89,CWKim-book,BGG-review-98}).
The
probability of $\nu_e\to\nu_{\mu,\tau}$ transitions
is easily obtained from the conservation of probability,
$
P^{\mathrm{Sun}}_{\nu_e\to\nu_{\mu,\tau}}
=
1
-
P^{\mathrm{Sun}}_{\nu_e\to\nu_e}
-
P^{\mathrm{Sun}}_{\nu_e\to\nu_s}
$,
yielding
\begin{equation}
P^{\mathrm{Sun}}_{\nu_e\to\nu_{\mu,\tau}}
=
\left( 1 - \cos^2{\vartheta_{23}} \cos^2{\vartheta_{24}} \right)
\left[
\frac{1}{2}
-
\left( \frac{1}{2} - P_c \right) \cos{2\vartheta_{12}} \, \cos{2\vartheta_{12}^M}(x_0)
\right]
\,.
\label{Pemt}
\end{equation}
Figure \ref{prob-sun} shows the probabilities
(\ref{Pee})--(\ref{Pemt})
for neutrinos produced in the center of the Sun
as functions of $p/\Delta_{21}$,
with
$ \sin^2 2\vartheta_{12} = 6 \times 10^{-3} $.
The dashed, solid and dotted curves
correspond to
$ \cos^2{\vartheta_{23}} \cos^2{\vartheta_{24}} = 0 $
($\nu_e\to\nu_{\mu,\tau}$ transitions only),
$ \cos^2{\vartheta_{23}} \cos^2{\vartheta_{24}} = 0.5 $
and
$ \cos^2{\vartheta_{23}} \cos^2{\vartheta_{24}} = 1 $
($\nu_e\to\nu_s$ transitions only),
respectively.
We have used the solar density distribution given in
\cite{Bahcall-WWW}.
From Fig. \ref{prob-sun}a
one can see that the survival probability of solar $\nu_e$'s
is sensitive to the value of
$ \cos^2{\vartheta_{23}} \cos^2{\vartheta_{24}} $
only in the range
$ 4 \times 10^4
\lesssim p/\Delta_{21} \lesssim
10^5 \, \mathrm{MeV}/\mathrm{eV}^2
$,
in which the resonance region is near the center
of the Sun and the shift of the resonance towards higher matter densities
for increasing values of
$ \cos^2{\vartheta_{23}} \cos^2{\vartheta_{24}} $
increases the survival probability.

Let us note that the transition probabilities of solar neutrinos can
be calculated more accurately by integrating
the evolution equation (\ref{evolution2})
along a realistic solar density distribution
\cite{Bahcall-WWW}.
Since the evolution of the amplitudes
$\psi^M_3=\psi'_3=\psi_3$
and
$\psi^M_4=\psi'_4=\psi_4$
is decoupled,
it is sufficient to integrate the two-dimensional
evolution equation
\begin{equation}
i \, \frac{ \mathrm{d} }{ \mathrm{d} x }
\left(
\begin{array}{c}
\psi'_1
\\
\psi'_2
\end{array}
\right)
=
\frac{1}{4p}
\left(
\begin{array}{cc}
\scriptstyle
- \Delta_{21} c_{2\vartheta_{12}} + 2 A_{CC}
&
\scriptstyle
\Delta_{21} s_{2\vartheta_{12}}
\\
\scriptstyle
\Delta_{21} s_{2\vartheta_{12}}
&
\scriptstyle
\Delta_{21} c_{2\vartheta_{12}}
- 2 c^2_{\vartheta_{23}} c^2_{\vartheta_{24}} A_{NC}
\end{array}
\right)
\left(
\begin{array}{c}
\psi'_1
\\
\psi'_2
\end{array}
\right)
\,,
\label{evolution3}
\end{equation}
where we have subtracted a common phase generated by $\Sigma_{21}$
that has no effect on flavor transitions.

It is important to note that the
evolution equation (\ref{evolution3})
and the equivalent probabilities
(\ref{Pee}), (\ref{Pes}) and (\ref{Pemt})
depend only on three parameters,
$\Delta_{21}$, $\sin{2\vartheta_{12}}$
and
$\cos^2{\vartheta_{23}} \cos^2{\vartheta_{24}}$,
\textit{i.e.} just one more than in the case of two generations
which is usually assumed in the analysis of solar neutrino data
(see
\cite{analysis-sun}
and references therein).
The two parameters
$\Delta_{21}$ and $\sin{2\vartheta_{12}}$
are analogous to the usual two-generation mixing parameters
$\Delta{m}^2$ and $\sin^2 2\vartheta$,
whereas
$\cos^2{\vartheta_{23}} \cos^2{\vartheta_{24}}$
is a new parameter characteristic of four-neutrino mixing.
The existence of only one additional parameter with respect to the
two-generation case implies that the analysis of
the solar neutrino data in the four-neutrino scheme
is feasible
(solar neutrino data have already been analyzed in
\cite{Fogli-Lisi-sun}
in the framework of three-neutrino mixing
in which there is also
one additional parameter compared to the
two-generation case).

\section{Regeneration of solar \lowercase{$\nu_e$'s} in the Earth}
\label{Regeneration}

The regeneration of solar electron neutrinos in the Earth
has been studied in several papers in the two-generation framework
(see
\cite{Mikheev-Smirnov-uspekhi-87,Baltz-Weneser-earth-87,%
Baltz-Weneser-earth-94,Lisi-Montanino-earth-97,%
Liu-Maris-Petcov-earth1-97,%
Petcov-diffractive-98,Chizhov-Maris-Petcov-98,%
Akhmedov-parametric-99,Chizhov-Petcov-earth-99,%
Dighe-Liu-Smirnov-earth-99,Guth-Randall-Serna-earth-99}
and references therein).
In this case,
the survival probability of solar $\nu_e$'s after passing
through the Earth is given by the standard formula
\cite{Mikheev-Smirnov-uspekhi-87,%
Baltz-Weneser-earth-87}
\begin{equation}
P^{\mathrm{Sun+Earth}}_{\nu_e\to\nu_e}
=
P^{\mathrm{Sun}}_{\nu_e\to\nu_e}
+
\frac{
\left( 1 - 2 P^{\mathrm{Sun}}_{\nu_e\to\nu_e} \right)
\left( P^{\mathrm{Earth}}_{\nu_2\to\nu_e} - \sin^2{\vartheta} \right)
}
{ \cos{2\vartheta} }
\,,
\label{P2SEee}
\end{equation}
where
$\vartheta$ is the two-generation mixing angle and
$P^{\mathrm{Earth}}_{\nu_2\to\nu_e}$
is the  probability of transitions
from the vacuum mass eigenstate $\nu_2$
to $\nu_e$ in the Earth.
One can see that if the matter effect inside the Earth is negligible,
$ P^{\mathrm{Earth}}_{\nu_2\to\nu_e} $
is given by
$|U_{e2}|^2=\sin^2{\vartheta}$
and Eq. (\ref{P2SEee})
reduces to
$
P^{\mathrm{Sun+Earth}}_{\nu_e\to\nu_e}
=
P^{\mathrm{Sun}}_{\nu_e\to\nu_e}
$.

Let us calculate the expression for the probability
$P^{\mathrm{Sun+Earth}}_{\nu_e\to\nu_\alpha}$
of $\nu_e\to\nu_\alpha$ transitions
($\alpha=e,s,\mu,\tau$)
of the solar neutrinos after passing
through the Earth
in the case of four-neutrino mixing.
We consider different final flavors $\alpha$
because we are interested not only in the
survival probability of electron neutrinos,
but also in the probabilities
of $\nu_e\to\nu_s$ and $\nu_e\to\nu_{\mu,\tau}$ transitions.
In the four-neutrino scheme under consideration
the state describing a neutrino reaching the Earth from the Sun
is a superposition of the two light vacuum mass eigenstates
$\nu_1$ and $\nu_2$:
\begin{equation}
|\nu_{\mathrm{Sun}}\rangle
=
\psi_1 \, |\nu_1\rangle
+
\psi_2 \, |\nu_2\rangle
\,,
\label{nuS}
\end{equation}
with the amplitudes $\psi_1$ and $\psi_2$
given by Eq. (\ref{final-amplitude1}).
These amplitudes are connected to the
average probability of $\nu_e\to\nu_\alpha$ transitions in the Sun
(as those given in Eqs. (\ref{Pee}) and (\ref{Pes}))
by
\begin{equation}
|\psi_1|^2
=
1 - |\psi_2|^2
=
\frac
{ P^{\mathrm{Sun}}_{\nu_e\to\nu_\alpha} - |U_{\alpha2}|^2 }
{ |U_{\alpha1}|^2 - |U_{\alpha2}|^2 }
\,.
\label{psi}
\end{equation}
Passing through the Earth, the state (\ref{nuS})
evolves to
\begin{equation}
|\nu_{\mathrm{Sun+Earth}}\rangle
=
\mathcal{T} \, |\nu_{\mathrm{Sun}}\rangle
=
\psi_1 \, \mathcal{T} \, |\nu_1\rangle
+
\psi_2 \, \mathcal{T} \, |\nu_2\rangle
\,,
\label{nuSE}
\end{equation}
where the operator $\mathcal{T}$
describes the evolution inside the Earth,
including the transitions between the states
$|\nu_1\rangle$
and
$|\nu_2\rangle$
induced by the matter effects.
The state $|\nu_{\mathrm{Sun+Earth}}\rangle$ describes the neutrinos
reaching the detector.
The probability of $\nu_e\to\nu_\alpha$ transitions
after crossing the Earth is given by
\begin{equation}
P^{\mathrm{Sun+Earth}}_{\nu_e\to\nu_\alpha}
=
|\langle\nu_\alpha|\nu_{\mathrm{Sun+Earth}}\rangle|^2
=
\left|
\psi_1 \langle \nu_\alpha|\mathcal{T}|\nu_1\rangle
+
\psi_2 \langle \nu_\alpha|\mathcal{T}|\nu_2\rangle
\right|^2
\,.
\label{PSEealpha1}
\end{equation}
Since the interference terms are not observable
\cite{Dighe-Liu-Smirnov-earth-99},
the measurable average transition probability is
\begin{equation}
P^{\mathrm{Sun+Earth}}_{\nu_e\to\nu_\alpha}
=
|\psi_1|^2
\,
P^{\mathrm{Earth}}_{\nu_1\to\nu_\alpha}
+
|\psi_2|^2
\,
P^{\mathrm{Earth}}_{\nu_2\to\nu_\alpha}
\,,
\label{PSEealpha2}
\end{equation}
where
$
P^{\mathrm{Earth}}_{\nu_k\to\nu_\alpha}
=
\left| \langle \nu_\alpha|\mathcal{T}|\nu_k\rangle \right|^2
$
is the probability of
$\nu_k\to\nu_\alpha$
transitions in the Earth.
Since the probabilities
$
P^{\mathrm{Earth}}_{\nu_1\to\nu_\alpha}
$
and
$
P^{\mathrm{Earth}}_{\nu_2\to\nu_\alpha}
$
are connected by the unitarity relation
\begin{equation}
P^{\mathrm{Earth}}_{\nu_1\to\nu_\alpha}
+
P^{\mathrm{Earth}}_{\nu_2\to\nu_\alpha}
+
|U_{\alpha3}|^2
+
|U_{\alpha4}|^2
=
1
\,,
\label{unitarity}
\end{equation}
using Eq. (\ref{psi}),
the probability of $\nu_e\to\nu_\alpha$ transitions
after crossing the Earth (\ref{PSEealpha1})
can be written as
\begin{equation}
P^{\mathrm{Sun+Earth}}_{\nu_e\to\nu_\alpha}
=
P^{\mathrm{Sun}}_{\nu_e\to\nu_\alpha}
+
\frac{
\left(
|U_{\alpha1}|^2 + |U_{\alpha2}|^2 - 2 P^{\mathrm{Sun}}_{\nu_e\to\nu_\alpha}
\right)
\left( P^{\mathrm{Earth}}_{\nu_2\to\nu_\alpha} - |U_{\alpha2}|^2 \right)
}
{ |U_{\alpha1}|^2 - |U_{\alpha2}|^2 }
\,.
\label{PSEealpha3}
\end{equation}
This equation is a four-neutrino analogy
of the two-generation formula (\ref{P2SEee}).
Again,
one can see that if the matter effects inside the Earth are negligible,
$ P^{\mathrm{Earth}}_{\nu_2\to\nu_\alpha} = |U_{\alpha2}|^2 $
and Eq. (\ref{PSEealpha3})
reduces to
$
P^{\mathrm{Sun+Earth}}_{\nu_e\to\nu_\alpha}
=
P^{\mathrm{Sun}}_{\nu_e\to\nu_\alpha}
$.
Moreover,
the expression in (\ref{PSEealpha3}) for
$P^{\mathrm{Sun+Earth}}_{\nu_e\to\nu_\alpha}$
is valid in schemes with any number of neutrinos
larger than four
in which only the smallest mass-squared difference contributes
to solar neutrino oscillations
(rearranging the mass eigenstate indices,
if necessary).
From Eq. (\ref{PSEealpha3}),
for
the probabilities of $\nu_e\to\nu_e$ and $\nu_e\to\nu_s$ transitions
after crossing the Earth we obtain
\begin{equation}
P^{\mathrm{Sun+Earth}}_{\nu_e\to\nu_e}
=
P^{\mathrm{Sun}}_{\nu_e\to\nu_e}
+
\frac{
\left(
1 - 2 P^{\mathrm{Sun}}_{\nu_e\to\nu_e}
\right)
\left( P^{\mathrm{Earth}}_{\nu_2\to\nu_e} - \sin^2{\vartheta_{12}} \right)
}
{ \cos{2\vartheta_{12}} }
\label{PSEee}
\end{equation}
and
\begin{equation}
P^{\mathrm{Sun+Earth}}_{\nu_e\to\nu_s}
=
P^{\mathrm{Sun}}_{\nu_e\to\nu_s}
+
\frac{
\left(
2 P^{\mathrm{Sun}}_{\nu_e\to\nu_s}
-
\cos^2{\vartheta_{23}} \cos^2{\vartheta_{24}}
\right)
\left(
P^{\mathrm{Earth}}_{\nu_2\to\nu_s}
-
\cos^2{\vartheta_{12}} \cos^2{\vartheta_{23}} \cos^2{\vartheta_{24}}
\right)
}
{ \cos{2\vartheta_{12}} \cos^2{\vartheta_{23}} \cos^2{\vartheta_{24}} }
\,,
\label{PSEes}
\end{equation}
with the probabilities
$P^{\mathrm{Sun}}_{\nu_e\to\nu_e}$
and
$P^{\mathrm{Sun}}_{\nu_e\to\nu_s}$
given by Eqs. (\ref{Pee}) and (\ref{Pes}),
respectively.
The
probability of $\nu_e\to\nu_{\mu,\tau}$ transitions
can easily be obtained from the conservation of probability,
$
P^{\mathrm{Sun+Earth}}_{\nu_e\to\nu_{\mu,\tau}}
=
1
-
P^{\mathrm{Sun+Earth}}_{\nu_e\to\nu_e}
-
P^{\mathrm{Sun+Earth}}_{\nu_e\to\nu_s}
$.
One can see that the $\nu_e$ survival probability
(\ref{PSEee})
has the same form as in the two-generation case (see Eq. (\ref{P2SEee})).

The expressions (\ref{PSEee}) and (\ref{PSEes})
are not defined for $\cos{2\vartheta_{12}}=0$.
In this case
$ P^{\mathrm{Sun}}_{\nu_e\to\nu_e} = 1 / 2 $
and
$
P^{\mathrm{Sun}}_{\nu_e\to\nu_s}
=
\cos^2{\vartheta_{23}} \cos^2{\vartheta_{24}} / 2
$,
as can be seen from Eqs. (\ref{Pee}) and (\ref{Pes}),
independently of the values of $\psi_1$ and $\psi_2$
in Eq. (\ref{nuS}).
In this case,
Eq. (\ref{psi}) does not apply
and
one must calculate $|\psi_1|^2=1-|\psi_2|^2$
using Eq. (\ref{final-amplitude1})
(or solving numerically the evolution equation (\ref{evolution3}) in the Sun)
and
determine $P^{\mathrm{Sun+Earth}}_{\nu_e\to\nu_\alpha}$
from Eqs. (\ref{PSEealpha2}) and (\ref{unitarity}).

The expression (\ref{PSEes}) is also not well defined for
$\cos^2{\vartheta_{23}} \cos^2{\vartheta_{24}} = 0$.
This is due to the fact that in this case one has
$
P^{\mathrm{Sun}}_{\nu_e\to\nu_s}
=
0
$,
as can be seen from Eq. (\ref{Pes}),
independently of the values of $\psi_1$ and $\psi_2$
in Eq. (\ref{nuS}),
and thus
Eq. (\ref{psi}) does not apply.
Since $\cos^2{\vartheta_{23}} \cos^2{\vartheta_{24}} = 0$
implies that
$
|U_{s3}|^2
+
|U_{s4}|^2
=
1
$,
the unitarity relation (\ref{unitarity})
leads to
$
P^{\mathrm{Earth}}_{\nu_1\to\nu_s}
=
P^{\mathrm{Earth}}_{\nu_2\to\nu_s}
=
0
$
and
we have
$ P^{\mathrm{Sun+Earth}}_{\nu_e\to\nu_s} = 0 $
from Eq. (\ref{PSEealpha2}).

The probability $P^{\mathrm{Earth}}_{\nu_2\to\nu_\alpha}$
can be calculated by integrating numerically
the evolution equation (\ref{evolution3}) in the Earth
with the initial amplitudes at the point $y_0$
of neutrinos entering the Earth
\begin{equation}
\left(
\begin{array}{c}
\psi'_1(y_0)
\\
\psi'_2(y_0)
\end{array}
\right)
=
\left(
\begin{array}{c}
\sin{\vartheta_{12}}
\\
\cos{\vartheta_{12}}
\end{array}
\right)
\,.
\label{initial-amplitude-earth}
\end{equation}
The probability $P^{\mathrm{Earth}}_{\nu_2\to\nu_\alpha}(y)$
at the point $y$ is given by
\begin{equation}
P^{\mathrm{Earth}}_{\nu_2\to\nu_\alpha}(y)
=
|\psi_\alpha(y)|^2
=
\left| \sum_{k=1,2} U'_{{\alpha}k} \, \psi'_k(y) \right|^2
\,.
\label{final-prob-earth}
\end{equation}

Another method to calculate
the probability $P^{\mathrm{Earth}}_{\nu_2\to\nu_\alpha}$
is based on the
approximation of the internal composition of the Earth
with two or more shells of density
(the minimum two shells correspond to the core and the mantle)
\cite{GKM-atm-98,%
Petcov-diffractive-98,Chizhov-Maris-Petcov-98,%
Akhmedov-parametric-99,Chizhov-Petcov-earth-99,%
Akhmedov-Dighe-Lipari-Smirnov-parametric-99}.
In each shell the corresponding energy eigenstates
evolve adiabatically, developing a phase,
and the amplitudes in confining shells are matched on the shell boundaries
imposing the continuity of the flavor amplitudes.
The initial amplitudes in the mass basis are
\begin{equation}
\Psi^0(y_0)
=
(0,1,0,0)^T
\,,
\label{initial-earth-mass}
\end{equation}
corresponding to $|\nu_2\rangle$.
The final amplitude in the rotated basis
\begin{equation}
\Psi'(y_n)=(\psi'_1(y_n),\psi'_2(y_n),\psi'_3(y_n),\psi'_4(y_n))^T
\,.
\label{final-earth1}
\end{equation}
after crossing the boundaries of
$n$ slabs of matter with constant density
at the points $y_1$, $y_2$, \ldots, $y_n$
is
\begin{eqnarray}
\Psi'(y_n)
&=&
\left[
U^M_{12} \Phi(y_n-y_{n-1}) {U^M_{12}}^\dagger
\right]_{(n)}
\left[
U^M_{12} \Phi(y_{n-1}-y_{n-2}) {U^M_{12}}^\dagger
\right]_{(n-1)}
\cdots
\nonumber
\\
&&
\cdots
\left[
U^M_{12} \Phi(y_2-y_1) {U^M_{12}}^\dagger
\right]_{(2)}
\left[
U^M_{12} \Phi(y_1-y_0) {U^M_{12}}^\dagger
\right]_{(1)}
U_{12} \, \Psi^0(y_0)
\,,
\label{final-earth2}
\end{eqnarray}
where
the notation $[\ldots]_{(i)}$
indicates that the matter-dependent quantities inside
of the square brackets must be evaluated with the matter density
in the $i^{\mathrm{th}}$ slab that extends from
$y_{i-1}$ to $y_i$.
The quantities
$\Phi(y)$
are the diagonal matrices of phases acquired by the
energy eigenstates in matter:
\begin{equation}
\Phi(y)
=
\mathrm{diag}\!\left(
e^{-i E^M_1 y},
e^{-i E^M_2 y},
e^{-i E^M_3 y},
e^{-i E^M_4 y}
\right)
\,.
\label{phases1}
\end{equation}
However,
because of the structure of the initial amplitudes (\ref{initial-earth-mass})
and the fact that at each boundary between two slabs
only $U^M_{12}$ operates,
the phases of the two decoupled energy eigenstates
$|\nu_3\rangle$,
$|\nu_4\rangle$
never come into play in Eq. (\ref{final-earth2}).
Therefore,
only the 1--2 sector is relevant.
Defining
\begin{equation}
\widetilde\Psi'(y_n)
=
\left(
\begin{array}{c}
\psi'_1(y_n)
\\
\psi'_2(y_n)
\end{array}
\right)
\,,
\quad
\widetilde\Psi^0(y_0)
=
\left(
\begin{array}{c}
\psi_1(y_0)
\\
\psi_2(y_0)
\end{array}
\right)
=
\left(
\begin{array}{c}
0
\\
1
\end{array}
\right)
\,,
\quad
\widetilde\Phi(y)
=
\mathrm{diag}\!\left(
e^{-i E^M_1 y},
e^{-i E^M_2 y}
\right)
\,,
\label{tilde}
\end{equation}
we have
\begin{eqnarray}
\widetilde\Psi'(y_n)
&=&
\left[
\widetilde{U}^M_{12}
\widetilde\Phi(y_n-y_{n-1})
\left.\widetilde{U}^M_{12}\right.^\dagger
\right]_{(n)}
\left[
\widetilde{U}^M_{12}
\widetilde\Phi(y_{n-1}-y_{n-2})
\left.\widetilde{U}^M_{12}\right.^\dagger
\right]_{(n-1)}
\cdots
\nonumber
\\
&&
\cdots
\left[
\widetilde{U}^M_{12} \widetilde\Phi(y_2-y_1)
\left.\widetilde{U}^M_{12}\right.^\dagger
\right]_{(2)}
\left[
\widetilde{U}^M_{12} \widetilde\Phi(y_1-y_0)
\left.\widetilde{U}^M_{12}\right.^\dagger
\right]_{(1)}
{\widetilde{U}}_{12} \, \widetilde\Psi^0(y_0)
\,,
\label{final-earth3}
\end{eqnarray}
where
$\widetilde{U}_{12}$
and
$\widetilde{U}^M_{12}$
are $2{\times}2$
matrices obtained from the 1--2 sectors of
$U_{12}$
and
$U^M_{12}$,
respectively.
Moreover,
since the probability of $\nu_2\to\nu_\alpha$
transitions in the Earth
is given by
\begin{equation}
P^{\mathrm{Earth}}_{\nu_2\to\nu_\alpha}
=
|\psi_\alpha(y_n)|^2
=
\left|
\sum_{k=1,2}
U'_{{\alpha}k} \, \psi'_k(y_n)
\right|^2
\,,
\label{PE2e}
\end{equation}
one can eliminate a common phase
in Eq. (\ref{final-earth3}).
Eliminating the phase generated by the lightest energy eigenstate,
the effective matrices of phases $\widetilde\Phi(y)$
to be used in Eq. (\ref{final-earth3})
for practical calculations are
\begin{equation}
\widetilde\Phi(y)
=
\mathrm{diag}\!\left(
1,
e^{-i\Delta^M_{21}y/2p}
\right)
\,,
\label{phases2}
\end{equation}
with
\begin{equation}
\Delta^M_{21}
=
2 p \left( E^M_2 - E^M_1 \right)
=
\sqrt{
\left( \Delta_{21} \cos{2\vartheta_{12}} - A \right)^2
+
\left( \Delta_{21} \sin{2\vartheta_{12}} \right)^2
}
\,.
\label{DeltaM21}
\end{equation}

The formalism presented in this section
and
in the previous one is appropriate for
analyzing the solar neutrino data in the framework of the four-neutrino
scheme B (see Eq. (\ref{AB}))
and in the scheme A with the exchange $ 1,2 \leftrightarrows 3,4 $
of the mass eigenstate indices.
Given the correspondence
of $\Delta_{21}$ and $\sin^2{2\vartheta_{12}}$
with the usual two-generation mixing parameters
$\Delta{m}^2$ and $\sin^2 2\vartheta$,
the oscillations of the solar neutrinos,
including matter effects,
in the four-neutrino schemes
depend on only one additional quantity,
$ \cos^2{2\vartheta_{23}} \cos^2{2\vartheta_{24}} $,
with respect to the two-generation case.
Therefore,
we believe that the analysis of the solar neutrino data
in a four-neutrino framework is well within
the reach of the groups specializing in this task.

\section{Atmospheric neutrinos}
\label{Atmospheric neutrinos}

The flavor evolution of the atmospheric neutrinos while
crossing the Earth depends on matter
\cite{atm-matter,Fogli-Lisi-atm,Narayan-three-nu,GKM-atm-98,%
Akhmedov-Dighe-Lipari-Smirnov-parametric-99}
if the matter-induced potential
is of the same order or larger than
the energy difference
$\Delta_{43}/2p$
of the two mass eigenstates $\nu_3$, $\nu_4$,
responsible for the atmospheric neutrino oscillations.
Since the electron and neutron number densities in the Earth
are approximately equal and vary from about
$ 2 \, N_A \mathrm{cm}^{-3} $ in the crust to about
$ 6 \, N_A \mathrm{cm}^{-3} $ in the center,
one can see
from Eqs. (\ref{dm2-atm}) and (\ref{VCC})
that matter effects are important for atmospheric neutrinos
with energy larger than about 10 GeV.

A convenient parameterization
of the mixing matrix $U$
for studying the matter effects for the atmospheric neutrinos is
\begin{equation}
U = V_{24} \, V_{23} \, V_{14} \, V_{13} \, V_{34} \, V_{12}
\,,
\label{U-atm}
\end{equation}
with
\begin{equation}
(V_{ij})_{ab}
=
\delta_{ab}
+
\left( \cos{\varphi_{ij}} - 1 \right)
\left( \delta_{ia} \delta_{ib} + \delta_{ja} \delta_{jb} \right)
+
\sin{\varphi_{ij}}
\left( \delta_{ia} \delta_{jb} - \delta_{ja} \delta_{ib} \right)
\,.
\label{atm-rotation}
\end{equation}
Here we have changed the notation for the rotation matrices
from
$U_{ij}$
to
$V_{ij}$
and the notation for the mixing angles
from
$\vartheta_{ij}$
to
$\varphi_{ij}$,
in order to emphasize that the
parameterization of $U$
is different from that used in Sections \ref{Solar neutrinos}
and
\ref{Regeneration}
for the solar neutrinos.

It is convenient to work in the rotated basis
\begin{equation}
\Psi'
=
{V'}^\dagger \, \Psi
\,,
\label{atm-rotated-basis-1}
\end{equation}
with
\begin{equation}
V'
=
V_{24} \, V_{23} \, V_{14} \, V_{13}
\,.
\label{atm-Up}
\end{equation}
In this case
the relation between the rotated basis (\ref{atm-rotated-basis-1})
and the mass basis (\ref{mass-basis}) is
\begin{equation}
\Psi'
=
V_{34} \, V_{12} \, \Psi^0
\,,
\label{atm-rotated-basis-2}
\end{equation}
and the evolution equation for $\Psi'$ is given by Eq. (\ref{evolution2}),
with the effective Hamiltonian
\begin{equation}
\mathcal{H}'
=
\frac{1}{2p}
\left( V_{34} \, V_{12} \, \mathcal{M}_0^2 \, V_{12}^\dagger \, V_{34}^\dagger
+ {V'}^\dagger \, \mathcal{A} \, V' \right)
\,.
\label{atm-hamiltonian2}
\end{equation}

As in the case of the solar neutrinos,
the results of the Bugey $\bar\nu_e$ disappearance experiment
\cite{Bugey-95}
imply the constraint (\ref{small}),
\textit{i.e.}
$
\sin^2{\varphi_{13}} + \sin^2{\varphi_{14}} \lesssim 10^{-2}
$.
Hence, we can neglect the two small angles
$\varphi_{13}$
and
$\varphi_{14}$
whose contributions to
the atmospheric neutrino oscillations are negligible.
In this case,
the explicit form of the mixing matrix
$ U = V_{24} \, V_{23} \, V_{34} \, V_{12} $
is
\begin{equation}
U
=
\left(
\begin{array}{cccc} \scriptstyle
c_{\varphi_{12}}
& \scriptstyle
s_{\varphi_{12}}
& \scriptstyle
0
& \scriptstyle
0
\\ \scriptstyle
- s_{\varphi_{12}} c_{\varphi_{23}} c_{\varphi_{24}}
& \scriptstyle
c_{\varphi_{12}} c_{\varphi_{23}} c_{\varphi_{24}}
& \scriptstyle
s_{\varphi_{23}} c_{\varphi_{24}} c_{\varphi_{34}}
-
s_{\varphi_{24}} s_{\varphi_{34}}
& \scriptstyle
s_{\varphi_{23}} c_{\varphi_{24}} s_{\varphi_{34}}
+
s_{\varphi_{24}} c_{\varphi_{34}}
\\ \scriptstyle
s_{\varphi_{12}} s_{\varphi_{23}}
& \scriptstyle
- c_{\varphi_{12}} s_{\varphi_{23}}
& \scriptstyle
c_{\varphi_{23}} c_{\varphi_{34}}
& \scriptstyle
c_{\varphi_{23}} s_{\varphi_{34}}
\\ \scriptstyle
s_{\varphi_{12}} c_{\varphi_{23}} s_{\varphi_{24}}
& \scriptstyle
- c_{\varphi_{12}} c_{\varphi_{23}} s_{\varphi_{24}}
& \scriptstyle
- s_{\varphi_{23}} s_{\varphi_{24}} c_{\varphi_{34}}
- c_{\varphi_{24}} s_{\varphi_{34}}
& \scriptstyle
- s_{\varphi_{23}} s_{\varphi_{24}} s_{\varphi_{34}}
+ c_{\varphi_{24}} c_{\varphi_{34}}
\end{array}
\right)
\,.
\label{U-atm-explicit}
\end{equation}
Comparing with the explicit form
(\ref{U-sun-explicit})
of the parameterization of the mixing matrix
used in Sections \ref{Solar neutrinos}
and
\ref{Regeneration},
one can see that
$\varphi_{12}=\vartheta_{12}$,
but the other mixing angles are different.
The mixing matrix (\ref{U-atm-explicit})
entails the following limiting cases:
\begin{description}
\item[$\varphi_{23}=\varphi_{24}=0.$]
The mixing matrix $U$ assumes a block-diagonal form
in which
$\nu_e$, $\nu_s$
are linear combinations of
$\nu_1$, $\nu_2$
alone
and
$\nu_\mu$, $\nu_\tau$
are linear combinations of
$\nu_3$, $\nu_4$
alone.
Therefore, this case corresponds to pure
$\nu_e\to\nu_s$ transitions for the solar neutrinos
and pure
$\nu_\mu\to\nu_\tau$ transitions for the atmospheric neutrinos.
\item[$\varphi_{23}=\pi/2.$]
This case is excluded because it yields
$U_{\mu3}=U_{\mu4}=0$,
implying that
$\nu_\mu$
does not mix with the two massive neutrinos $\nu_3$, $\nu_4$.
Since the mass-squared difference
$m_4^2-m_3^2$
is responsible for atmospheric neutrino
oscillations,
the disappearance of the atmospheric $\nu_\mu$'s
cannot be explained.
\item[$\varphi_{24}=\pi/2.$]
In this case
$U_{s1}=U_{s2}=0$
and
the solar neutrinos undergo
$\nu_e\to\nu_\mu$ and/or $\nu_e\to\nu_\tau$ transitions,
whereas
the atmospheric neutrinos undergo
$\nu_\mu\to\nu_s$ and/or $\nu_\mu\to\nu_\tau$ transitions.
\item[$\varphi_{24}=\pi/2, \, \varphi_{23}=0.$]
The mixing matrix $U$ assumes a block-diagonal form
in which
$\nu_e$, $\nu_\tau$
are linear combinations of
$\nu_1$, $\nu_2$
alone
and
$\nu_\mu$, $\nu_s$
are linear combinations of
$\nu_3$, $\nu_4$
alone.
This case corresponds to pure
$\nu_e\to\nu_\tau$ transitions for the solar neutrinos
and pure
$\nu_\mu\to\nu_s$ transitions for the atmospheric neutrinos.
\end{description}

In the following we present a formalism
for the analysis of the oscillations of
the atmospheric neutrinos in the Earth
without making any assumption
about the value of $\varphi_{23}$,
but it is clear that in order to explain the disappearance of atmospheric
$\nu_\mu$'s observed in atmospheric neutrino experiments
the mixing angle $\varphi_{23}$
cannot be too large.
Indeed,
the up-down asymmetry of high-energy $\mu$-like events observed in the
Super-Kamiokande experiment \cite{SK-atm} implies
the upper bound \cite{BGGS-AB-99}
\begin{equation}
\sin^2{\varphi_{23}} = |U_{\mu1}|^2+|U_{\mu2}|^2 \lesssim 0.55
\,.
\label{s23-bound-1}
\end{equation}
Moreover,
the negative results of the short-baseline
CDHS and CCFR accelerator $\nu_\mu$ disappearance experiments
\cite{CDHS-84,CCFR-84}
imply that \cite{BGKP-96,BGG-AB-96}
\begin{equation}
\sin^2{\varphi_{23}} = |U_{\mu1}|^2+|U_{\mu2}|^2 \lesssim 0.2
\qquad \mbox{for} \qquad
\Delta{m}^2_{\mathrm{LSND}} \gtrsim 0.4 \, \mathrm{eV}^2
\,,
\label{s23-bound-2}
\end{equation}
and the limit is much more stringent for some
larger values of $\Delta{m}^2_{\mathrm{LSND}}$
(for example,
$ |U_{\mu1}|^2+|U_{\mu2}|^2 \lesssim 2 \times 10^{-2} $
for
$\Delta{m}^2_{\mathrm{LSND}} \simeq 2 \, \mathrm{eV}^2$,
see \cite{BGG-review-98}).
This limit is valid in a large part of the LSND-allowed range for
$\Delta{m}^2_{\mathrm{LSND}}=\Delta{m}^2_{41}$,
\begin{equation}
0.2 \, \mathrm{eV}^2
\lesssim
\Delta{m}^2_{\mathrm{LSND}}
\lesssim
2 \, \mathrm{eV}^2
\,.
\label{LSND-range}
\end{equation}

With the approximation
$\varphi_{13}=\varphi_{14}=0$,
the matrix $V'$ is given by
\begin{equation}
V'
=
V_{24} \, V_{23}
=
\left(
\begin{array}{cccc} \scriptstyle
1
& \scriptstyle
0
& \scriptstyle
0
& \scriptstyle
0
\\ \scriptstyle
0
& \scriptstyle
c_{\varphi_{23}} c_{\varphi_{24}}
& \scriptstyle
s_{\varphi_{23}} c_{\varphi_{24}}
& \scriptstyle
s_{\varphi_{24}}
\\ \scriptstyle
0
& \scriptstyle
- s_{\varphi_{23}}
& \scriptstyle
c_{\varphi_{23}}
& \scriptstyle
0
\\ \scriptstyle
0
& \scriptstyle
- c_{\varphi_{23}} s_{\varphi_{24}}
& \scriptstyle
- s_{\varphi_{23}} s_{\varphi_{24}}
& \scriptstyle
c_{\varphi_{24}}
\end{array}
\right)
\,,
\label{Ups-atm}
\end{equation}
and
the effective Hamiltonian $\mathcal{H}'$ is
\begin{equation}
\mathcal{H}'
=
\frac{1}{2p}
\left(
V_{34} \, V_{12} \, \mathcal{M}_0^2 \, V_{12}^\dagger \, V_{34}^\dagger
+
V_{23}^\dagger \, V_{24}^\dagger \, \mathcal{A} \, V_{24} \, V_{23}
\right)
\,,
\label{atm-hamiltonian3}
\end{equation}
which reads explicitly as
\begin{equation}
\mathcal{H}'
=
\frac{1}{4p}
\left(
\begin{array}{cccc}
\scriptstyle
\begin{array}{c}
\scriptstyle
\Sigma_{21} - \Delta_{21} c_{2\varphi_{12}}
\\
\scriptstyle
+ 2 A_{CC}
\end{array}
&
\scriptstyle
\Delta_{21} s_{2\varphi_{12}}
&
\scriptstyle
0
&
\scriptstyle
0
\\
\scriptstyle
\Delta_{21} s_{2\varphi_{12}}
&
\scriptstyle
\begin{array}{c}
\scriptstyle
\Sigma_{21} + \Delta_{21} c_{2\varphi_{12}}
\\
\scriptstyle
- 2 c^2_{\varphi_{23}} c^2_{\varphi_{24}} A_{NC}
\end{array}
&
\scriptstyle
- s_{2\varphi_{23}} c^2_{\varphi_{24}} A_{NC}
&
\scriptstyle
- c_{\varphi_{23}} s_{2\varphi_{24}} A_{NC}
\\
\scriptstyle
0
&
\scriptstyle
- s_{2\varphi_{23}} c^2_{\varphi_{24}} A_{NC}
&
\scriptstyle
\begin{array}{c}
\scriptstyle
\Sigma_{43} - \Delta_{43} c_{2\varphi_{34}}
\\
\scriptstyle
- 2 s^2_{\varphi_{23}} c^2_{\varphi_{24}} A_{NC}
\end{array}
&
\scriptstyle
\Delta_{43} s_{2\varphi_{34}}
- s_{\varphi_{23}} s_{2\varphi_{24}} A_{NC}
\\
\scriptstyle
0
&
\scriptstyle
- c_{\varphi_{23}} s_{2\varphi_{24}} A_{NC}
&
\scriptstyle
\Delta_{43} s_{2\varphi_{34}}
- s_{\varphi_{23}} s_{2\varphi_{24}} A_{NC}
&
\scriptstyle
\begin{array}{c}
\scriptstyle
\Sigma_{43} + \Delta_{43} c_{2\varphi_{34}}
\\
\scriptstyle
- 2 s^2_{\varphi_{24}} A_{NC}
\end{array}
\end{array}
\right)
\,,
\label{atm-hamiltonian4}
\end{equation}
with
$ \Sigma_{43} \equiv m_4^2 + m_3^2 $.

For high-energy atmospheric neutrinos crossing the Earth
$
\Sigma_{21} \sim \Delta_{21}
\ll
|A_{CC}| \sim |A_{NC}| \sim \Delta_{43}
\ll
\Sigma_{43} \sim 2 \Delta_{41}
$
and there is a resonance in the 3--4 sector of the effective
Hamiltonian (\ref{atm-hamiltonian4}) for
\begin{equation}
\left(
\sin^2{\varphi_{24}} - \sin^2{\varphi_{23}} \, \cos^2{\varphi_{24}}
\right)
A_{NC}
=
\Delta_{43} \, \cos{2\varphi_{34}}
\,.
\label{res-atm}
\end{equation}
Since $A_{NC}<0$,
there can be a resonance for neutrinos if
\begin{equation}
|\tan \varphi_{24}|
<
|\sin \varphi_{23}|
\qquad \mbox{and} \qquad
\cos{2\varphi_{34}} > 0
\label{res-atm-cond-1}
\end{equation}
or
\begin{equation}
|\tan \varphi_{24}|
>
|\sin \varphi_{23}|
\qquad \mbox{and} \qquad
\cos{2\varphi_{34}} < 0
\,.
\label{res-atm-cond-2}
\end{equation}
Otherwise,
there can be a resonance for antineutrinos,
for which $A_{NC}\to\overline{A}_{NC}>0$.
One can see that if
$ \cos{\varphi_{23}} = \cos{\varphi_{24}} = 1 $,
corresponding to pure atmospheric
$\nu_\mu\to\nu_\tau$
transitions,
the term in parenthesis in Eq. (\ref{res-atm}) vanishes
and
there is no resonance for any value of $A_{NC}$,
for both neutrinos and antineutrinos.
Indeed,
if
$ \cos{\varphi_{23}} = \cos{\varphi_{24}} = 1 $
the matter induced terms in the 3--4 sector
of the effective Hamiltonian (\ref{atm-hamiltonian4})
vanish and there is no matter effect
in atmospheric neutrino oscillations,
as expected from the correspondence of this case
with two-generation $\nu_\mu\to\nu_\tau$
transitions that are not affected by matter effects
because $\nu_\mu$ and $\nu_\tau$
feel the same matter potential.

\begin{figure}[t]
\begin{center}
\includegraphics[bb=81 554 447 768,width=0.80\linewidth]{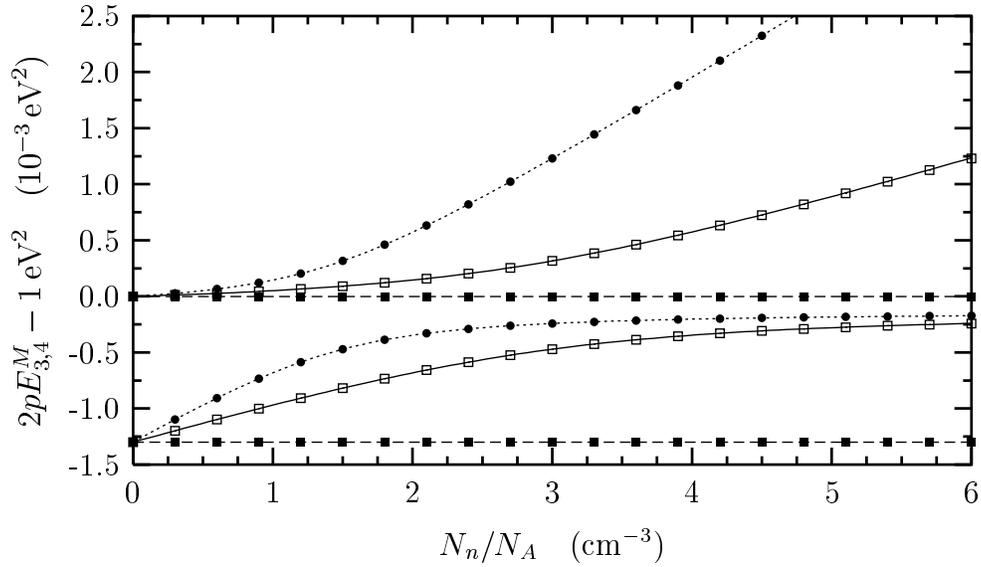}
\caption{ \label{eigen-atm}
The lines show the third and fourth eigenvalues
of the effective Hamiltonian (\ref{atm-hamiltonian4})
as functions of the neutron number density $N_n$
obtained with a numerical diagonalization,
whereas
the points show the eigenvalues obtained with Eq. (\ref{eigen-atm-34}).
We assumed
$ m_1 = 0 $,
$ \Delta_{21} = 1.0 \times 10^{-5} \, \mathrm{eV}^2 $,
$ \Delta_{43} = 1.3 \times 10^{-3} \, \mathrm{eV}^2 $,
$ \Delta_{41} = 1 \, \mathrm{eV}^2 $,
$ \sin^2{2\varphi_{12}} = 10^{-2} $,
$ \cos{\varphi_{23}} = 1 $,
$ \cos^2{\varphi_{34}} = 0.1 $,
$ p = 10 \, \mathrm{GeV} $
and
$ N_n = N_e $.
The dashed, solid and dotted lines
correspond, respectively, to
$
\cos^2{\varphi_{24}}
=
1, \,
0.5, \,
0
$.
}
\end{center}
\end{figure}

Since
$
\Sigma_{43} \sim 2 \Delta_{41}
\gg
\Delta_{43} \sim |A_{CC}| \sim |A_{NC}|
\gg
\Sigma_{21} \sim \Delta_{21}
$,
the evolution of the amplitudes
$\psi'_1$ and $\psi'_2$ is adiabatic and decoupled from the
evolution of the amplitudes
$\psi'_3$ and $\psi'_4$,
that are coupled by the matter effects.
To lowest order in the power expansion in the small quantities
$
|A_{CC}|/\Sigma_{43}
\sim
|A_{NC}|/\Sigma_{43}
\sim
\Delta_{43}/\Sigma_{43}
$
and neglecting
$\Sigma_{21} \sim \Delta_{21}$,
the eigenvalues of the effective Hamiltonian $\mathcal{H}'$ are
\begin{eqnarray}
E^M_1
&=&
- \cos^2{\varphi_{23}} \cos^2{\varphi_{24}} A_{NC} / 2 p
=
- \cos^2{\varphi_{23}} \cos^2{\varphi_{24}} V_{NC}
\,,
\label{eigen-atm-1}
\\
E^M_2
&=&
A_{CC} / 2 p
=
V_{CC}
\,,
\label{eigen-atm-2}
\\
E^M_{4,3}
&=&
\frac{1}{4p}
\left\{
\vphantom{\frac{1}{4p}}
\Sigma_{43}
-
\left(
s^2_{\varphi_{24}} + s^2_{\varphi_{23}} \, c^2_{\varphi_{24}}
\right)
A_{NC}
\right.
\nonumber
\\
&&
\left.
\vphantom{\frac{1}{4p}}
\pm
\sqrt{
\left[
\Delta_{43} \, c_{2\varphi_{34}}
-
\left(
s^2_{\varphi_{24}} - s^2_{\varphi_{23}} \, c^2_{\varphi_{24}}
\right)
A_{NC}
\right]^2
+
\left[
\Delta_{43} s_{2\varphi_{34}}
- s_{\varphi_{23}} s_{2\varphi_{24}} A_{NC}
\right]^2
}
\right\}
\,.
\label{eigen-atm-34}
\end{eqnarray}
Note that the two light eigenvalues
$E^M_1$ and $E^M_2$
depend on the matter potentials
$V_{NC}$ and $V_{CC}$,
respectively,
but are independent of the neutrino momentum $p$.
The accuracy of the expression (\ref{eigen-atm-34})
for the two heavy eigenvalues
$E^M_3$ and $E^M_4$
is shown in Fig. \ref{eigen-atm},
in which the lines show the third and fourth eigenvalues
of the effective Hamiltonian
$\mathcal{H}'$
as functions of the neutron number density $N_n$
obtained with a numerical diagonalization
and the points along the lines
show the eigenvalues obtained with Eq. (\ref{eigen-atm-34}).
The values of the mixing parameters used for drawing Fig. \ref{eigen-atm}
are
$ m_1 = 0 $,
$ \Delta_{21} = 1.0 \times 10^{-5} \, \mathrm{eV}^2 $,
$ \Delta_{43} = 1.3 \times 10^{-3} \, \mathrm{eV}^2 $,
$ \Delta_{41} = 1 \, \mathrm{eV}^2 $,
$ \sin^2{2\varphi_{12}} = 10^{-2} $,
$ \cos{\varphi_{23}} = 1 $,
$ \cos^2{\varphi_{34}} = 0.1 $.
We considered
$ p = 10 \, \mathrm{GeV} $
and
$ N_n = N_e $.
The dashed, solid and dotted lines
in Fig. \ref{eigen-atm}
correspond, respectively, to
$
\cos^2{\varphi_{24}}
=
1, \,
0.5, \,
0
$.
One can see that for
$\cos^2{\varphi_{24}}=1$
(dashed lines)
there is no matter effect.
Indeed,
since we assumed
$\cos{\varphi_{23}}=1$,
this case corresponds to
pure $\nu_\mu\to\nu_\tau$ transitions.
From Fig. \ref{eigen-atm}
one can also see that the resonance is more pronounced when
$\cos^2{\varphi_{24}}$ decreases,
in agreement with the condition (\ref{res-atm-cond-2})
and the fact that the case
$\cos^2{\varphi_{24}}=0$
(dotted lines)
corresponds to pure $\nu_\mu\to\nu_s$ transitions,
for which the effect of the
matter-induced neutral current potential is maximal.

As in the case of solar neutrinos discussed in Section \ref{Solar neutrinos},
apart from the common term $p+V_{NC}$
that has been subtracted from the beginning,
the eigenvalues $E^M_k$ ($k=1,2,3,4$)
are the energies of the energy eigenstates in matter
$|\nu^M_k\rangle$,
whose column matrix of amplitudes
$\Psi^M$
is related to $\Psi'$ by
\begin{equation}
\Psi^M
=
{V^M_{12}}^\dagger \, {V^M_{34}}^\dagger \, \Psi'
\,,
\label{atm-matter-mixing1}
\end{equation}
where
\begin{equation}
(V^M_{ij})_{ab}
=
\delta_{ab}
+
\left( \cos{\varphi_{ij}^M} - 1 \right)
\left( \delta_{ia} \delta_{ib} + \delta_{ja} \delta_{jb} \right)
+
\sin{\varphi_{ij}^M}
\left( \delta_{ia} \delta_{jb} - \delta_{ja} \delta_{ib} \right)
\qquad
(ij=12,34)
\label{VijM}
\end{equation}
are the orthogonal matrices that diagonalize the
1--2 and 3--4 sectors
of the effective Hamiltonian 
$\mathcal{H}'$ in Eq. (\ref{atm-hamiltonian4}).
The value of the effective mixing angle in matter $\varphi_{12}^M$
is given by
\begin{equation}
\tan 2\varphi_{12}^M
=
\frac
{ \tan 2\varphi_{12} }
{ 1 - A / \Delta_{21} \cos{2\varphi_{12}} }
\,.
\label{phi12m-1}
\end{equation}
However,
since $A \gg \Delta_{21}$,
we have
\begin{equation}
\varphi_{12}^M
\simeq
\pi/2
\,.
\label{phi12m-2}
\end{equation}
Therefore,
the mixing in the 1--2 sector is strongly suppressed.
The value of the effective mixing angle in matter $\varphi_{34}^M$
is given by
\begin{equation}
\tan 2\varphi^M_{34}
=
\frac
{
\Delta_{43} \sin{2\varphi_{34}}
- \sin{\varphi_{23}} \sin{2\varphi_{24}} A_{NC}
}
{
\Delta_{43} \, \cos{2\varphi_{34}}
-
\left(
\sin^2{\varphi_{24}} - \sin^2{\varphi_{23}} \, \cos^2{\varphi_{24}}
\right)
A_{NC}
}
\,.
\label{phi34m}
\end{equation}
When the resonance condition (\ref{res-atm})
is fulfilled,
the effective mixing angle $\varphi^M_{34}$
is equal to $\pi/4$,
\textit{i.e.} the mixing generated by $\varphi^M_{34}$
is maximal.
On the other hand,
when the equality
$
\Delta_{43} \sin{2\varphi_{34}}
=
\sin{\varphi_{23}} \sin{2\varphi_{24}} A_{NC}
$
is satisfied,
the mixing is suppressed,
as expected from the vanishing of the off-diagonal terms
in the 3--4 sector of the effective Hamiltonian 
$\mathcal{H}'$,
leading to $\varphi^M_{34}=0$ if
$
\Delta_{43} \, \cos{2\varphi_{34}}
>
\left(
\sin^2{\varphi_{24}} - \sin^2{\varphi_{23}} \, \cos^2{\varphi_{24}}
\right)
A_{NC}
$
and
$\varphi^M_{34}=\pi/2$ in the opposite case.
From Eq. (\ref{phi34m})
it is clear that if
$\varphi_{23}=\varphi_{24}=0$,
\textit{i.e.}
there are only $\nu_\mu\to\nu_\tau$ transitions,
we have
$\varphi^M_{34}=\varphi_{34}$,
because there is no matter effect,
and
if
$\varphi_{23}=0$,
$\varphi_{24}=\pi/2$,
\textit{i.e.}
there are only $\nu_\mu\to\nu_s$ transitions,
we have
\begin{equation}
\tan 2\varphi^M_{34}
=
\frac
{
\Delta_{43} \sin{2\varphi_{34}}
}
{
\Delta_{43} \, \cos{2\varphi_{34}}
-
A_{NC}
}
\,,
\label{phi34m-numu-nus}
\end{equation}
which is the usual expression for the effective mixing angle
in matter in two-generation
$\nu_\mu$--$\nu_s$ mixing.

The connection between the flavor basis
$\Psi$
and the energy basis in matter $\Psi^M$ is given by
\begin{equation}
\Psi
=
V' \, V^M_{34} \, V^M_{12} \, \Psi^M
=
U^M \, \Psi^M
\,,
\label{atm-matter-mixing2}
\end{equation}
where $ U^M = V' \, V^M_{34} \, V^M_{12} $
is the effective mixing matrix in matter
which can be obtained from the vacuum mixing matrix
(\ref{U-atm-explicit})
with the replacements
$\varphi_{12}\mapsto\pi/2$
and
$\varphi_{34}\mapsto\varphi_{34}^M$
(the mixing angles $\varphi_{23}$ and $\varphi_{24}$
are not modified in matter):
\begin{equation}
U^M
=
\left(
\begin{array}{cccc} \scriptstyle
0
& \scriptstyle
1
& \scriptstyle
0
& \scriptstyle
0
\\ \scriptstyle
- c_{\varphi_{23}} c_{\varphi_{24}}
& \scriptstyle
0
& \scriptstyle
s_{\varphi_{23}} c_{\varphi_{24}} c_{\varphi_{34}^M}
-
s_{\varphi_{24}} s_{\varphi_{34}^M}
& \scriptstyle
s_{\varphi_{23}} c_{\varphi_{24}} s_{\varphi_{34}^M}
+
s_{\varphi_{24}} c_{\varphi_{34}^M}
\\ \scriptstyle
s_{\varphi_{23}}
& \scriptstyle
0
& \scriptstyle
c_{\varphi_{23}} c_{\varphi_{34}^M}
& \scriptstyle
c_{\varphi_{23}} s_{\varphi_{34}^M}
\\ \scriptstyle
c_{\varphi_{23}} s_{\varphi_{24}}
& \scriptstyle
0
& \scriptstyle
- s_{\varphi_{23}} s_{\varphi_{24}} c_{\varphi_{34}^M}
- c_{\varphi_{24}} s_{\varphi_{34}^M}
& \scriptstyle
- s_{\varphi_{23}} s_{\varphi_{24}} s_{\varphi_{34}^M}
+ c_{\varphi_{24}} c_{\varphi_{34}^M}
\end{array}
\right)
\,.
\label{U-atm-matter-explicit}
\end{equation}

The amplitudes of the two light mass eigenstates
are related to the amplitudes of the flavor eigenstates by the relations
\begin{eqnarray}
\psi^M_1
& = &
- c_{\varphi_{23}} c_{\varphi_{24}} \psi_s
+ s_{\varphi_{23}} \psi_\mu
+ c_{\varphi_{23}} s_{\varphi_{24}} \psi_\tau
\,.
\label{psiM1}
\\
\psi^M_2
& = &
\psi_e
\,,
\label{psiM2}
\end{eqnarray}
that are independent of the matter density.
Furthermore,
Eqs. (\ref{atm-matter-mixing1}) and (\ref{phi12m-2}) imply that
$\psi^M_1=-\psi'_2$
and
$\psi^M_2=\psi'_1$,
independently of the value of the matter density.
Therefore,
the evolution of the amplitudes
$\psi^M_1$ and $\psi^M_2$
is adiabatic and decoupled from the
evolution of the amplitudes
$\psi^M_3$
and
$\psi^M_4$
that are coupled by the matter effects.
Indeed,
the amplitudes
$\psi^M_3$
and
$\psi^M_4$
are related to the amplitudes
$\psi'_3$
and
$\psi'_4$
by the relations
$
\psi^M_3
=
\cos{\varphi_{34}^M} \psi'_3
-
\sin{\varphi_{34}^M} \psi'_4
$
and
$
\psi^M_4
=
\sin{\varphi_{34}^M} \psi'_3
+
\cos{\varphi_{34}^M} \psi'_4
$,
with the mixing angle $\varphi_{34}^M$
which depends on the matter density.
This means that
$ \nu^M_3 \leftrightarrows \nu^M_4 $
transitions occur when the variation of the matter density is very rapid.

If a flavor neutrino $\nu_\alpha$ has been produced
at the initial coordinate $z_0$
along the neutrino trajectory,
we have
\begin{equation}
\psi^{M(\alpha)}_k(z_0)
=
\left. U^M_{{\alpha}k} \right|_{z_0}
\qquad
(k=1,2,3,4)
\,,
\label{atm-initial}
\end{equation}
where we have introduced the superscript
$(\alpha)$
to indicate explicitly the initial neutrino flavor.
The corresponding amplitudes
$\psi^{M(\alpha)}_1(z)$
and
$\psi^{M(\alpha)}_2(z)$
at the coordinate $z$ are given by
\begin{equation}
\psi^{M(\alpha)}_k(z)
=
\exp\left( -i \int_{z_0}^{z} E^M_k \, \mathrm{d}z \right)
U^M_{{\alpha}k}
\qquad
(k=1,2)
\,,
\label{psiM12}
\end{equation}
where the integral must be evaluated along the neutrino path.
The elements $U^M_{{\alpha}1}$ and $U^M_{{\alpha}2}$
of the effective mixing matrix in matter (\ref{U-atm-matter-explicit})
do not carry the subscript $z_0$
because they are independent of the matter density,
as long as $A \gg \Delta_{21}$
and Eq. (\ref{phi12m-2}) is satisfied.

Because of the decoupling of the evolution of the amplitudes
$\psi'_1=\psi^M_2$
and
$\psi'_2=-\psi^M_1$,
the evolution in matter of the amplitudes
$\psi'_3$ and $\psi'_4$
can be calculated by integrating numerically
the two-dimensional evolution equation
\begin{equation}
i \, \frac{ \mathrm{d} }{ \mathrm{d} x }
\left(
\begin{array}{c}
\psi''_3
\\
\psi''_4
\end{array}
\right)
=
\frac{1}{4p}
\left(
\begin{array}{cc}
\scriptstyle
- \Delta_{43} c_{2\varphi_{34}}
- 2 s^2_{\varphi_{23}} c^2_{\varphi_{24}} A_{NC}
&
\scriptstyle
\Delta_{43} s_{2\varphi_{34}}
- s_{\varphi_{23}} s_{2\varphi_{24}} A_{NC}
\\
\scriptstyle
\Delta_{43} s_{2\varphi_{34}}
- s_{\varphi_{23}} s_{2\varphi_{24}} A_{NC}
&
\scriptstyle
\Delta_{43} c_{2\varphi_{34}}
- 2 s^2_{\varphi_{24}} A_{NC}
\end{array}
\right)
\left(
\begin{array}{c}
\psi''_3
\\
\psi''_4
\end{array}
\right)
\,,
\label{atm-evolution3}
\end{equation}
obtained from the 3--4 sector of (\ref{atm-hamiltonian4})
and the definition
\begin{equation}
\psi'_k(z)
=
e^{-i\Sigma_{43}(z-z_0)} \, \psi''_k(z)
\qquad
(k=3,4)
\,,
\label{psipp}
\end{equation}
that eliminates
the large common phase generated by $\Sigma_{43}$.
For an initial flavor neutrino $\nu_\alpha$,
the amplitudes in the rotated basis at the coordinate $z_0$
of neutrino production in the atmosphere are given by
\begin{equation}
\psi^{\prime\prime(\alpha)}_k(z_0)
=
\psi^{\prime(\alpha)}_k(z_0)
=
V'_{{\alpha}k}
\qquad
(k=3,4)
\,,
\label{initial-amplitude-atm-1}
\end{equation}
and we use the notation
$\psi^{\prime(\alpha)}_k(z)$,
$\psi^{\prime\prime(\alpha)}_k(z)$
for the the amplitudes in the rotated basis
at the coordinate $z$ along the neutrino path
obtained by solving the differential equations
(\ref{atm-evolution3})
with the initial conditions (\ref{initial-amplitude-atm-1}).
For example,
for an initial muon neutrino we have
\begin{equation}
\left(
\begin{array}{c}
\psi^{\prime\prime(\mu)}_3(z_0)
\\
\psi^{\prime\prime(\mu)}_4(z_0)
\end{array}
\right)
=
\left(
\begin{array}{c}
\psi^{\prime(\mu)}_3(z_0)
\\
\psi^{\prime(\mu)}_4(z_0)
\end{array}
\right)
=
\left(
\begin{array}{c}
\cos{\varphi_{23}}
\\
0
\end{array}
\right)
\,,
\label{initial-amplitude-atm-3}
\end{equation}
to be used as the initial condition
for the numerical solution of the differential equations (\ref{atm-evolution3}).
The corresponding initial amplitudes
$\psi^{M(\mu)}_1(z_0)$ and $\psi^{M(\mu)}_2(z_0)$
in Eq.(\ref{psiM12}) are
\begin{equation}
\psi^{M(\mu)}_1(z_0)
=
U^M_{\mu1}
=
\sin{\varphi_{23}}
\,,
\qquad
\psi^{M(\mu)}_2(z_0)
=
U^M_{\mu2}
=
0
\,.
\label{initial-amplitude-atm-4}
\end{equation}

The probability $P^{\mathrm{Atm}}_{\nu_\alpha\to\nu_\beta}(z)$
of
$\nu_\alpha\to\nu_\beta$
transitions
at the point $z$ is given by
\begin{equation}
P^{\mathrm{Atm}}_{\nu_\alpha\to\nu_\beta}(z)
=
|\psi^{(\alpha)}_\beta(z)|^2
=
\left| \sum_{k=1}^4 V'_{{\beta}k} \, \psi^{\prime(\alpha)}_k(z) \right|^2
\,,
\label{final-prob-atm-1}
\end{equation}
where
$\psi^{(\alpha)}_\beta(z)$
indicates the amplitude of the flavor state $|\nu_\beta\rangle$
at the coordinate $z$ along the neutrino path
obtained from a pure flavor state
$|\nu_\alpha\rangle$
at the initial coordinate $z_0$.

Since in real experiments
the interference terms depending on the large phases
$e^{-i\Sigma_{43}(z-z_0)}$
are washed out by the averages over the energy resolution of the detector
and over the production region,
the measurable probability is given by
\begin{equation}
P^{\mathrm{Atm}}_{\nu_\alpha\to\nu_\beta}(z)
=
\left| \sum_{k=1,2} V'_{{\beta}k} \, \psi^{\prime(\alpha)}_k(z) \right|^2
+
\left| \sum_{k=3,4} V'_{{\beta}k} \, \psi^{\prime\prime(\alpha)}_k(z) \right|^2
\,,
\label{final-prob-atm-2}
\end{equation}
which can be written as
\begin{equation}
P^{\mathrm{Atm}}_{\nu_\alpha\to\nu_\beta}(z)
=
\left| \sum_{k=1,2} U^M_{{\beta}k} \, \psi^{M(\alpha)}_k(z) \right|^2
+
\left| \sum_{k=3,4} V'_{{\beta}k} \, \psi^{\prime\prime(\alpha)}_k(z) \right|^2
\,.
\label{final-prob-atm-3}
\end{equation}
Furthermore,
using Eq. (\ref{psiM12})
we obtain
\begin{equation}
P^{\mathrm{Atm}}_{\nu_\alpha\to\nu_\beta}(z)
=
\left|
U^M_{{\beta}1} U^M_{{\alpha}1}
+
U^M_{{\beta}2} U^M_{{\alpha}2}
\exp\left( -i \int_{z_0}^{z} \Delta{E}^M_{21} \, \mathrm{d}z \right)
\right|^2
+
\left| \sum_{k=3,4} V'_{{\beta}k} \, \psi^{\prime\prime(\alpha)}_k(z) \right|^2
\,,
\label{final-prob-atm-4}
\end{equation}
with
\begin{equation}
\Delta{E}^M_{21}
=
E^M_2 - E^M_1
=
V_{CC} + \cos^2{\varphi_{23}} \cos^2{\varphi_{24}} V_{NC}
\,.
\label{DeltaM21-atm}
\end{equation}

Since
$U^M_{\alpha1}=U^M_{\alpha1}(1-\delta_{{\alpha}e})$,
$U^M_{\alpha2}=\delta_{{\alpha}e}$
and
$V'_{e3}=V'_{e4}=0$,
from Eq. (\ref{final-prob-atm-4})
one can see that
the survival probability of atmospheric electron neutrinos
is one, independently of the matter effect
(this is a consequence of
the approximation
$U_{13}=U_{14}=0$):
\begin{equation}
P^{\mathrm{Atm}}_{\nu_e\to\nu_e} = 1
\,.
\label{Pee-atm}
\end{equation}
Therefore
\begin{equation}
P^{\mathrm{Atm}}_{\nu_e\to\nu_\beta}
=
P^{\mathrm{Atm}}_{\nu_\beta\to\nu_e}
=
0
\qquad
(\beta\not=e)
\,.
\label{Pebeta}
\end{equation}
The other transition probabilities are given by
\begin{equation}
P^{\mathrm{Atm}}_{\nu_\alpha\to\nu_\beta}(z)
=
|U^M_{{\beta}1}|^2 |U^M_{{\alpha}1}|^2
+
\left| \sum_{k=3,4} V'_{{\beta}k} \, \psi^{\prime\prime(\alpha)}_k(z) \right|^2
\qquad
(\alpha,\beta\not=e)
\,,
\label{final-prob-atm-5}
\end{equation}
with the amplitudes
$\psi^{\prime\prime(\alpha)}_3(z)$
and
$\psi^{\prime\prime(\alpha)}_4(z)$
obtained from the numerical solution of Eq. (\ref{atm-evolution3})
with the initial condition (\ref{initial-amplitude-atm-1}).
It is important to note that the elements
$U^M_{{\alpha}1}$
of the effective mixing matrix in matter (\ref{U-atm-matter-explicit})
depend only on the vacuum mixing angles
$\varphi_{23}$
and
$\varphi_{24}$
and therefore they are independent of the matter density.

The transition probabilities of atmospheric antineutrinos
can be obtained from Eq. (\ref{final-prob-atm-5})
changing the signs of the matter potentials.
Note that,
although
we neglected possible CP-violating phases in the neutrino mixing matrix,
the transition probabilities are not CP-symmetric,
\begin{equation}
P^{\mathrm{Atm}}_{\nu_\alpha\to\nu_\beta}
\not=
P^{\mathrm{Atm}}_{\bar\nu_\alpha\to\bar\nu_\beta}
\,,
\label{CP-asymm}
\end{equation}
because the matter effect is CP-asymmetric,
but they are T-symmetric,
\begin{equation}
P^{\mathrm{Atm}}_{\nu_\alpha\to\nu_\beta}
=
P^{\mathrm{Atm}}_{\nu_\beta\to\nu_\alpha}
\,,
\label{T-symm}
\end{equation}
because
the matter distribution is symmetric along the path
of neutrinos crossing the Earth.

Instead of solving numerically the evolution equation
(\ref{atm-evolution3}) for atmospheric neutrinos propagating in the Earth,
the amplitudes
$\psi^{\prime\prime(\alpha)}_3(z)$
and
$\psi^{\prime\prime(\alpha)}_4(z)$
can be calculated approximating the internal composition of the Earth
with two or more shells of constant density,
as done for the regeneration of solar $\nu_e$'s in Earth
in Section \ref{Regeneration}.
In this case,
after crossing the boundaries of
$n$ matter-slabs of constant density
at the points $z_1$, $z_2$, \ldots, $z_n$
we have
\begin{eqnarray}
\widetilde\Psi^{\prime\prime(\alpha)}(z_n)
&=&
\left[
\widetilde{V}^M_{34}
\widetilde\Phi(z_n-z_{n-1})
\left.\widetilde{V}^M_{34}\right.^\dagger
\right]_{(n)}
\left[
\widetilde{V}^M_{34}
\widetilde\Phi(z_{n-1}-z_{n-2})
\left.\widetilde{V}^M_{34}\right.^\dagger
\right]_{(n-1)}
\cdots
\nonumber
\\
&&
\cdots
\left[
\widetilde{V}^M_{34}
\widetilde\Phi(z_2-z_1)
\left.\widetilde{V}^M_{34}\right.^\dagger
\right]_{(2)}
\left[
\widetilde{V}^M_{34}
\widetilde\Phi(z_1-z_0)
\left.\widetilde{V}^M_{34}\right.^\dagger
\right]_{(1)}
\widetilde\Psi^{\prime\prime(\alpha)}(z_0)
\,,
\label{final-atm-2}
\end{eqnarray}
where
$\widetilde{V}^M_{34}$
is the $2{\times}2$
matrix obtained from the 3--4 sector of
$V^M_{34}$
and
\begin{equation}
\widetilde\Psi^{\prime\prime(\alpha)}(z)
=
\left(
\begin{array}{c}
\psi^{\prime\prime(\alpha)}_3(z)
\\
\psi^{\prime\prime(\alpha)}_4(z)
\end{array}
\right)
\,,
\quad
\widetilde\Phi(z)
=
\mathrm{diag}\!\left(
1,
e^{-i\Delta^M_{43}z/2p}
\right)
\,,
\label{atm-tilde}
\end{equation}
with
\begin{eqnarray}
\Delta^M_{43}
&=&
2 p \left( E^M_4 - E^M_3 \right)
\nonumber
\\
&=&
\sqrt{
\left[
\Delta_{43} \, c_{2\varphi_{34}}
-
\left(
s^2_{\varphi_{24}} - s^2_{\varphi_{23}} \, c^2_{\varphi_{24}}
\right)
A_{NC}
\right]^2
+
\left[
\Delta_{43} s_{2\varphi_{34}}
- s_{\varphi_{23}} s_{2\varphi_{24}} A_{NC}
\right]^2
}
\,.
\label{DeltaM43}
\end{eqnarray}

The formalism presented in this section 
can be applied to the analysis of the
 atmospheric neutrino data
in the framework of the four-neutrino
scheme B (see Eq. (\ref{AB}))
and in the scheme A with the exchange $ 1,2 \leftrightarrows 3,4 $
of the mass eigenstate indices.
Whereas in the case of solar neutrinos
the presence of only one additional parameter
($ \cos^2{\vartheta_{23}} \cos^2{\vartheta_{24}} $)
distinguishes the oscillation probabilities
in the framework of four-neutrino mixing
from those in two-neutrino mixing,
in the case of atmospheric neutrinos
there are two relevant mixing angles,
$\varphi_{23}$
and
$\varphi_{24}$,
in addition to the mixing parameters
$\Delta_{43}$ and $\varphi_{34}$
corresponding to the usual two-generation mixing parameters
$\Delta{m}^2$ and $\vartheta$.
However,
information on the values of these parameters
can be obtained from solar neutrino data
(as discussed above,
since the parameterizations of the mixing matrix $U$
that we have used for the study of solar and atmospheric neutrinos
are different,
we have only the equality
$\varphi_{21}=\vartheta_{21}$
and
a numerical transformation between the two parameterizations is needed
in practical calculations in order to connect the values of the other mixing
angles).
Therefore,
we think that the best way to obtain information on four-neutrino mixing
from the data of solar and atmospheric neutrino experiments
is to perform a combined fit.
This task is challenging,
but far from impossible
(analyses of the atmospheric neutrino data
in the framework of
three-neutrino mixing taking into account the matter effects
have been already carried out
\cite{Fogli-Lisi-atm,Narayan-three-nu,GKM-atm-98,%
Akhmedov-Dighe-Lipari-Smirnov-parametric-99}).

\section{Long-baseline experiments}
\label{Long-baseline experiments}

The formalism presented in the previous section
for studying the oscillations of the atmospheric neutrinos
with matter effects
can also be applied to calculate the oscillation probabilities
in long-baseline experiments (LBL) \cite{LBL}.
Indeed,
the region of the parameter space explored by long-baseline
neutrino oscillation experiments
is similar to that explored by atmospheric neutrino experiments.
Furthermore,
the neutrino beams in long-baseline experiments
propagate in the mantle of the Earth,
where the matter density
($\rho_{\mathrm{mantle}} \simeq 4.5 \, \mathrm{g}/\mathrm{cm}^3$)
and electron number fraction ($Y_e\simeq0.5$)
are approximately constant.
Therefore,
we can apply the formula in Eq. (\ref{final-atm-2})
with only one slab:
\begin{equation}
\widetilde\Psi^{\prime\prime(\alpha)}(z_1)
=
\widetilde{V}^M_{34}
\widetilde\Phi(z_1-z_0)
\left.\widetilde{V}^M_{34}\right.^\dagger
\widetilde\Psi^{\prime\prime(\alpha)}(z_0)
\,,
\label{LBL-ampli-1}
\end{equation}
with all the matter quantities evaluated for an approximate constant
density in the mantle.
Taking into account Eqs. (\ref{initial-amplitude-atm-1})
and
(\ref{atm-tilde}),
we obtain
\begin{equation}
\sum_{k=3,4} V'_{{\beta}k} \, \psi^{\prime\prime(\alpha)}_k(z_1)
=
U^M_{{\beta}3} U^M_{{\alpha}3}
+
U^M_{{\beta}4} U^M_{{\alpha}4}
\exp\left( -i \frac{\Delta^M_{43}(z_1-z_0)}{2p} \right)
\,.
\label{LBL-ampli-2}
\end{equation}
Inserting this expression
in Eq. (\ref{final-prob-atm-4}),
one obtains that
the probability of $\nu_\alpha\to\nu_\beta$ transitions
with $\alpha,\beta\not=e$
in long-baseline experiments
is given by
\begin{equation}
P^{\mathrm{LBL}}_{\nu_\alpha\to\nu_\beta}(L)
=
|U^M_{{\beta}1}|^2 |U^M_{{\alpha}1}|^2
+
\left|
U^M_{{\beta}3} U^M_{{\alpha}3}
+
U^M_{{\beta}4} U^M_{{\alpha}4}
\exp\left( -i \frac{\Delta^M_{43}L}{2p} \right)
\right|^2
\qquad
(\alpha,\beta\not=e)
\,,
\label{prob-LBL}
\end{equation}
where $L=z_1-z_0$
is the source-detector distance.
From the discussion in Section \ref{Atmospheric neutrinos}
it is clear that
$
P^{\mathrm{LBL}}_{\nu_e\to\nu_e}=1
$
and
$
P^{\mathrm{LBL}}_{\nu_e\to\nu_\beta}
=
P^{\mathrm{LBL}}_{\nu_\beta\to\nu_e}
=
0
$
for $\beta\not=e$.
Hence,
in the following part of this Section we restrict the values of
the flavor indices $\alpha$, $\beta$ to $s,\mu,\tau$.

The survival probability of $\nu_\alpha$,
can be written as
\begin{equation}
P^{\mathrm{LBL}}_{\nu_\alpha\to\nu_\alpha}(L)
=
|U^M_{{\alpha}1}|^4
+
\left( 1 - |U^M_{{\alpha}1}|^2 \right)^2
P^{\mathrm{LBL};3,4}_{\nu_\alpha\to\nu_\alpha}(L)
\,,
\label{prob-LBL-surv}
\end{equation}
with
\begin{equation}
P^{\mathrm{LBL};3,4}_{\nu_\alpha\to\nu_\alpha}(L)
=
1
-
\frac
{ 4 |U^M_{{\alpha}3}|^2 |U^M_{{\alpha}4}|^2 }
{ \left( |U^M_{{\alpha}3}|^2 + |U^M_{{\alpha}4}|^2 \right)^2 }
\,
\sin^2\left( \frac{\Delta^M_{43}L}{4p} \right)
\,.
\label{Psurv-34-1}
\end{equation}
Defining
\begin{equation}
\cos\xi_{\alpha}
=
\frac
{ U^M_{{\alpha}3} }
{ \sqrt{ |U^M_{{\alpha}3}|^2 + |U^M_{{\alpha}4}|^2 } }
\,,
\qquad
\sin\xi_{\alpha}
=
\frac
{ U^M_{{\alpha}4} }
{ \sqrt{ |U^M_{{\alpha}3}|^2 + |U^M_{{\alpha}4}|^2 } }
\,,
\label{cosxiLBL}
\end{equation}
Eq. (\ref{Psurv-34-1})
can be written as
\begin{equation}
P^{\mathrm{LBL};3,4}_{\nu_\alpha\to\nu_\alpha}(L)
=
1
-
\sin^2 2\xi_{\alpha}
\,
\sin^2\left( \frac{\Delta^M_{43}L}{4p} \right)
\,,
\label{Psurv-34-2}
\end{equation}
which shows that
$P^{\mathrm{LBL};3,4}_{\nu_\alpha\to\nu_\alpha}(L)$
is a two-generation-like
survival probability generated by the mixing angle
$\xi_{\alpha}$.
Since
$P^{\mathrm{LBL};3,4}_{\nu_\alpha\to\nu_\alpha}(L)$
is bounded by
\begin{equation}
\left(
\frac
{ |U^M_{{\alpha}3}|^2 - |U^M_{{\alpha}4}|^2 }
{ |U^M_{{\alpha}3}|^2 + |U^M_{{\alpha}4}|^2 }
\right)^2
=
1
-
\sin^2 2\xi_{\alpha}
\leq
P^{\mathrm{LBL};3,4}_{\nu_\alpha\to\nu_\alpha}(L)
\leq
1
\,,
\label{Psurv-34-bounds}
\end{equation}
the survival probability
(\ref{prob-LBL-surv})
is restricted to the range
\begin{equation}
|U^M_{{\alpha}1}|^4
+
\left( |U^M_{{\alpha}3}|^2 - |U^M_{{\alpha}4}|^2 \right)^2
\leq
P^{\mathrm{LBL}}_{\nu_\alpha\to\nu_\alpha}(L)
\leq
|U^M_{{\alpha}1}|^4
+
\left( 1 - |U^M_{{\alpha}1}|^2 \right)^2
\,.
\label{Psurv-bounds}
\end{equation}
In the limit
$U^M_{{\alpha}1}=0$
we have
$
P^{\mathrm{LBL}}_{\nu_\alpha\to\nu_\alpha}(L)
=
P^{\mathrm{LBL};3,4}_{\nu_\alpha\to\nu_\alpha}(L)
$,
\textit{i.e.}
the survival probability of long-baseline $\nu_\alpha$'s
reduces to the two-generation form
(\ref{Psurv-34-2}).

 Eqs. (\ref{prob-LBL-surv})
and (\ref{Psurv-34-2})
seem to depend on only three parameters,
$|U_{\alpha1}|^2$,
$\sin^2 2\xi_{\alpha}$
and
$\Delta^M_{43}$.
But since
$\sin^2 2\xi_{\alpha}$
and
$\Delta^M_{43}$
depend on the neutrino momentum $p$
through the matter-induced quantity $A_{CC}$,
the analysis of
long-baseline $\nu_\alpha$ disappearance experiments
requires one to express
$\sin^2 2\xi_{\alpha}$
and
$\Delta^M_{43}$
in terms of the mixing angles
$\varphi_{23}$,
$\varphi_{24}$,
$\varphi_{34}$
and
$\Delta_{43}$.
Because the matter density and composition are known and constant,
such an analysis is possible.

Of particular interest is
the survival probability of $\nu_\mu$'s
that constitute the main part
of the initial beam in long-baseline experiments.
It is given by
\begin{equation}
P^{\mathrm{LBL}}_{\nu_\alpha\to\nu_\alpha}(L)
=
\sin^4{\varphi_{23}}
+
\cos^4{\varphi_{23}}
\,
P^{\mathrm{LBL};3,4}_{\nu_\mu\to\nu_\mu}(L)
\,,
\label{Pmumu-LBL}
\end{equation}
where
$P^{\mathrm{LBL};3,4}_{\nu_\mu\to\nu_\mu}(L)$
is given by Eq. (\ref{Psurv-34-2})
with
\begin{equation}
\xi_\mu = \varphi^M_{34}
\,.
\label{ximu}
\end{equation}
If $\varphi_{23}=0$,
which is consistent with the experimental bounds
(\ref{s23-bound-1}) and (\ref{s23-bound-2}),
the survival probability (\ref{Pmumu-LBL})
reduces to the two-generation form
$P^{\mathrm{LBL};3,4}_{\nu_\mu\to\nu_\mu}(L)$.
However,
even if $\varphi_{23}=0$,
in general, there are simultaneous
$\nu_\mu\to\nu_\tau$
and
$\nu_\mu\to\nu_s$
transitions,
with pure
$\nu_\mu\to\nu_\tau$
transitions only if
$\varphi_{24}=0$
and pure
$\nu_\mu\to\nu_s$
transitions only if
$\varphi_{24}=\pi/2$.

The probabilities of $\nu_\alpha\to\nu_\beta$
transitions with $\alpha\neq\beta$ are given by
\begin{equation}
P^{\mathrm{LBL}}_{\nu_\alpha\to\nu_\beta}(L)
=
2 \, |U^M_{{\alpha}1}|^2 \, |U^M_{{\beta}1}|^2
-
4 \, U^M_{\alpha3} \, U^M_{\beta3} \, U^M_{\alpha4} \, U^M_{\beta4}
\,
\sin^2\left( \frac{\Delta^M_{43}L}{4p} \right)
\qquad
(\alpha\neq\beta)
\,.
\label{prob-LBL-trans}
\end{equation}
In particular,
the probabilities of
$\nu_\mu\to\nu_\tau$
and
$\nu_\mu\to\nu_s$
transitions are
\begin{eqnarray}
P^{\mathrm{LBL}}_{\nu_\mu\to\nu_\tau}(L)
&=&
\frac{1}{2} \, \sin^2{2\varphi_{23}} \sin^2{\varphi_{24}}
+
\cos^2{\varphi_{23}} \sin{2\varphi^M_{34}}
\left[
\left(
\cos^2{\varphi_{24}}
-
\sin^2{\varphi_{23}} \sin^2{\varphi_{24}}
\right)
\sin{2\varphi^M_{34}}
\right.
\nonumber
\\
&&
\hspace{2cm}
\left.
+
\sin{\varphi_{23}} \sin{2\varphi_{24}} \cos{2\varphi^M_{34}}
\right]
\sin^2\left( \frac{\Delta^M_{43}L}{4p} \right)
\,,
\label{Pmutau-LBL}
\\
P^{\mathrm{LBL}}_{\nu_\mu\to\nu_s}(L)
&=&
\frac{1}{2} \, \sin^2{2\varphi_{23}} \cos^2{\varphi_{24}}
+
\cos^2{\varphi_{23}} \sin{2\varphi^M_{34}}
\left[
\left(
\sin^2{\varphi_{24}}
-
\sin^2{\varphi_{23}} \cos^2{\varphi_{24}}
\right)
\sin{2\varphi^M_{34}}
\right.
\nonumber
\\
&&
\hspace{2cm}
\left.
-
\sin{\varphi_{23}} \sin{2\varphi_{24}} \cos{2\varphi^M_{34}}
\right]
\sin^2\left( \frac{\Delta^M_{43}L}{4p} \right)
\,.
\label{Pmus-LBL}
\end{eqnarray}

In the limit
$\varphi_{23}=0$
we have
\begin{eqnarray}
P^{\mathrm{LBL}}_{\nu_\mu\to\nu_\tau}(L)
&=&
\cos^2{\varphi_{24}}
\,
\sin^2{2\varphi^M_{34}}
\,
\sin^2\left( \frac{\Delta^M_{43}L}{4p} \right)
\,,
\label{Pmutau-LBL-1}
\\
P^{\mathrm{LBL}}_{\nu_\mu\to\nu_s}(L)
&=&
\sin^2{\varphi_{24}}
\,
\sin^2{2\varphi^M_{34}}
\,
\sin^2\left( \frac{\Delta^M_{43}L}{4p} \right)
\,,
\label{Pmus-LBL-1}
\end{eqnarray}
\textit{i.e.}
the
$\nu_\mu\to\nu_\tau$
and
$\nu_\mu\to\nu_s$
transition probabilities
are given by two-generation-like expressions
weighted by the factors
$\cos^2{\varphi_{24}}$
and
$\sin^2{\varphi_{24}}$,
respectively,
resulting in pure $\nu_\mu\to\nu_\tau$ oscillations
(with $\varphi^M_{34}=\varphi_{34}$ and $\Delta^M_{43}=\Delta_{43}$,
\textit{i.e.} without matter effects)
if
$\varphi_{24}=0$
and
pure $\nu_\mu\to\nu_s$ oscillations
(with
$\varphi^M_{34}$ given by Eq. (\ref{phi34m-numu-nus})
and
$
\Delta^M_{43}
=
\sqrt{
\left[
\Delta_{43} \, c_{2\varphi_{34}}
-
A_{NC}
\right]^2
+
\left[
\Delta_{43} s_{2\varphi_{34}}
\right]^2
}
$)
if
$\varphi_{24}=\pi/2$.

\section{Conclusions}
\label{Conclusions}

We have presented a formalism
which is appropriate for analyzing the results of
the solar, atmospheric and long-baseline experiments
in the framework of four-neutrino mixing with matter effects.
We have considered the two four-neutrino mixing schemes A and B
in Eq. (\ref{AB})
that are compatible with the results of all
the neutrino oscillation experiments
\cite{BGG-AB-96,Barger-variations-98,BGGS-AB-99}.
Only the scheme B has been considered explicitly here,
but all the results are also valid in the scheme A
and can be obtained with the exchanges $ 1,2 \leftrightarrows 3,4 $
of the mass eigenstate indices.
Since the elements $U_{e3}$ and $U_{e4}$
of the neutrino mixing matrix in the scheme B
are known to be small
(see Eq. (\ref{small}))
and negligible for the oscillations of
the solar and atmospheric neutrinos,
we have used the approximation
$U_{e3}=U_{e4}=0$.

For the solar neutrinos the matter effects are important in the case of
MSW transitions in the Sun
(Section \ref{Solar neutrinos})
and
for the regeneration of $\nu_e$'s in the Earth
(Section \ref{Regeneration}).
The oscillations of
the solar neutrinos depend only on three parameters,
the mass-squared difference $\Delta_{21}$,
the mixing angle $\vartheta_{12}$
and the product
$\cos^2{\vartheta_{23}} \cos^2{\vartheta_{24}}$
(the parameterization of the mixing matrix
for the study of solar neutrinos in terms of the mixing angles
$\vartheta_{12}$,
$\vartheta_{23}$,
$\vartheta_{24}$
and
$\vartheta_{34}$
is given explicitly in Eq. (\ref{U-sun-explicit})).
Therefore,
only one more parameter than in the case of two-neutrino mixing
is needed in order to analyze the results of
the solar neutrino experiments
in the framework of four-neutrino mixing.
Since
the solar neutrino data have already been analyzed
in the framework of three-neutrino mixing
\cite{Fogli-Lisi-sun}
in which there is also
one additional parameter compared to the
two-generation case,
the analysis of solar neutrino data
in the framework of four-neutrino mixing
using the formalism presented here
is a feasible task.

As described in Section \ref{Atmospheric neutrinos},
the oscillations of the atmospheric neutrinos in the Earth
depend on
the mass-squared difference $\Delta_{43}$
and the three mixing angles
$\varphi_{23}$,
$\varphi_{24}$
and
$\varphi_{34}$
(the parameterization of the mixing matrix
for the study of the atmospheric neutrinos
in terms of the mixing angles
$\varphi_{12}$,
$\varphi_{23}$,
$\varphi_{24}$
and
$\varphi_{34}$
is given explicitly
in Eq. (\ref{U-atm-explicit})).
Hence,
in the analysis of
the atmospheric neutrino data in the framework of four-neutrino
mixing
there are
two parameters more than in the case of two-neutrino mixing.
The best way to constrain these parameters
would be to perform a combined fit of the results of
the solar, atmospheric and possibly short-baseline
and long-baseline neutrino oscillation experiments.
The formalism presented in this paper can be used to perform this task.

In conclusion,
we would like to note that as a consequence of the approximation
$U_{e3}=U_{e4}=0$
the electron neutrinos do not oscillate in
the atmospheric and long-baseline
experiments.
Because of the upper bound
(\ref{small}),
this approximation is acceptable for
the solar and atmospheric experiments,
but may be inadequate in the case of high-precision long-baseline experiments
that could observe $\nu_\mu\to\nu_e$ transitions
below the bound implied by the constraint (\ref{small})
(see \cite{BGG-bounds-98}).
Furthermore,
we have neglected the possible presence of CP-violating phases in the mixing
matrix,
because CP violation is not observable in
the solar neutrino experiments.
We note that only the probability of
$\nu_e$ survival
and
$\nu_e\to\nu_{\mu,\tau}$ transitions
can be measured in the solar neutrino experiments.
In the atmospheric neutrino experiments
detected neutrinos and antineutrinos are not distinguished
and
the effects of the CP-violating phases are washed 
out by the average over the energy resolution of the detector
and over the neutrino path length.
However,
it is possible that
the effects of CP violation will be measured in
high-precision
long-baseline experiments
\cite{BGG-CP-98}
with conventional neutrino beams or with neutrino beams from
a neutrino factory
\cite{nufactory}.
In this case, an exact formalism of neutrino oscillations with
$U_{e3}\neq0$,
$U_{e4}\neq0$
and CP-violating phases in the neutrino mixing matrix
is necessary \cite{DGKK-preparation}.
\vspace{-.5 cm}
\acknowledgments
\vspace{-.5 cm}
Three of us (D.D., C.G., and K.K.)
would like to express their gratitude to
the Korea Institute for Advanced Study (KIAS)
for kind hospitality.
In addition,
D.D. wishes to thank the U.S. Department of Education for financial
support via the Graduate Assistance in Areas of National Need (GAANN)
program and
K.K. is supported in part by the U.S. Department 
of Energy Grant
DE-FG02-91ER40688-Task A.
Finally,
C.G. would like to thank
G.L. Fogli,
M.C. Gonzalez-Garcia,
P.I. Krastev, C.N. Leung,
A. Marchionni,
C. Pena-Garay, S.T. 
\nopagebreak[4]
Petcov, A. Romanino and J. Sato
for useful discussions at the
$\nu$-Fact'99 workshop in Lyon.


\begin{thebibliography}{10}

\bibitem{SK-atm}
Y. Fukuda \textit{et al.} (Super-Kamiokande Coll.), Phys. Rev. Lett.
  \textbf{81}, 1562 (1998); K. Scholberg (Super-Kamiokande Coll.),
  hep-ex/9905016.

\bibitem{exp-atm}
Y. Fukuda \textit{et al.}, Phys. Rev. Lett. \textbf{82}, 2644 (1999); A. Habig
  (Super-Kamiokande Coll.), hep-ex/9903047; Y. Fukuda \textit{et al.}
  (Kamiokande Coll.), Phys. Lett. \textbf{B335}, 237 (1994); R. Becker-Szendy
  \textit{et al.} (IMB Coll.), Nucl. Phys. B (Proc. Suppl.) \textbf{38}, 331
  (1995); W.W.M. Allison \textit{et al.} (Soudan Coll.), Phys. Lett.
  \textbf{B449}, 137 (1999); M. Ambrosio \textit{et al.} (MACRO Coll.), Phys.
  Lett. \textbf{B434}, 451 (1998).

\bibitem{exp-sun}
B.T. Cleveland \textit{et al.} (Homestake Coll.), Astrophys. J. \textbf{496},
  505 (1998); K.S. Hirata \textit{et al.} (Kamiokande Coll.), Phys. Rev. Lett.
  \textbf{77}, 1683 (1996); W. Hampel \textit{et al.} (GALLEX Coll.), Phys.
  Lett. \textbf{B447}, 127 (1999); J.N. Abdurashitov \textit{et al.} (SAGE
  Coll.), astro-ph/9907113; Y. Fukuda (Super-Kamiokande Coll.), Phys. Rev.
  Lett. \textbf{81}, 1158 (1998); Erratum \textit{ibid.} \textbf{81}, 4279
  (1998); 
  hep-ex/9812011; M.B. Smy (Super-Kamiokande Coll.), hep-ex/9903034.

\bibitem{LSND}
C. Athanassopoulos \textit{et al.} (LSND Coll.), Phys. Rev. Lett. \textbf{75},
  2650 (1995); Phys. Rev. Lett. \textbf{77}, 3082 (1996); Phys. Rev. Lett.
  \textbf{81}, 1774 (1998).

\bibitem{Giunti-baksan-99}
C. Giunti, hep-ph/9907485, Talk given at the X$^{\mathrm{th}}$ International
  Baksan School \textit{Particles and Cosmology}, Baksan Valley,
  Karbardino-Balkaria, Russia, 19--25 April 1999.

\bibitem{analysis-sun}
J.N. Bahcall, P.I. Krastev and A.Yu. Smirnov, Phys. Rev. \textbf{D58}, 096016
  (1998), hep-ph/9807216; Y. Fukuda \textit{et al.}, Phys. Rev. Lett.
  \textbf{82}, 1810 (1999), hep-ex/9812009; V. Barger and K. Whisnant, Phys.
  Lett. \textbf{B456}, 54 (1999), hep-ph/9903262; M.C. Gonzalez-Garcia
  \textit{et al.}, hep-ph/9906469.

\bibitem{MSW}
S.P. Mikheyev and A.Yu. Smirnov, Yad. Fiz. \textbf{42}, 1441 (1985) [Sov. J.
  Nucl. Phys. \textbf{42}, 913 (1985)]; Il Nuovo Cim. \textbf{C9}, 17 (1986);
  L. Wolfenstein, Phys. Rev. \textbf{D17}, 2369 (1978); Phys. Rev.
  \textbf{D20}, 2634 (1979).

\bibitem{four-models}
J.T. Peltoniemi, D. Tommasini and J.W.F. Valle, Phys. Lett. \textbf{B298}, 383
  (1993); E.J. Chun \textit{et al.}, Phys. Lett. \textbf{B357}, 608 (1995);
  S.C. Gibbons \textit{et al.}, Phys. Lett. \textbf{B430}, 296 (1998); B.
  Brahmachari and R.N. Mohapatra, Phys. Lett. \textbf{B437}, 100 (1998); S.
  Mohanty, D.P. Roy and U. Sarkar, Phys. Lett. \textbf{B445}, 185 (1998); J.T.
  Peltoniemi and J.W.F. Valle, Nucl. Phys. \textbf{B406}, 409 (1993); Q.Y. Liu
  and A.Yu. Smirnov, Nucl. Phys. \textbf{B524}, 505 (1998); D.O. Caldwell and
  R.N. Mohapatra, Phys. Rev. \textbf{D48}, 3259 (1993); E. Ma and P. Roy, Phys.
  Rev. \textbf{D52}, R4780 (1995); A.Yu. Smirnov and M. Tanimoto, Phys. Rev.
  \textbf{D55}, 1665 (1997); N. Gaur \textit{et al.}, Phys. Rev. \textbf{D58},
  071301 (1998); E.J. Chun, C.W. Kim and U.W. Lee, Phys. Rev. \textbf{D58},
  093003 (1998); K. Benakli and A.Yu. Smirnov, Phys. Rev. Lett. \textbf{79},
  4314 (1997); Y. Chikira, N. Haba and Y. Mimura, hep-ph/9808254; C. Liu and J.
  Song, hep-ph/9812381; W. Grimus, R. Pfeiffer and T. Schwetz, hep-ph/9905320.

\bibitem{four-phenomenology}
J.J. Gomez-Cadenas and M.C. Gonzalez-Garcia, Z. Phys. \textbf{C71}, 443 (1996),
  hep-ph/9504246; S. Goswami, Phys. Rev. \textbf{D55}, 2931 (1997),
  hep-ph/9507212; V. Barger, Y.B. Dai, K. Whisnant and B.L. Young, Phys. Rev.
  \textbf{D59}, 113010 (1999), hep-ph/9901388; V. Barger, T.J. Weiler and K.
  Whisnant, Phys. Lett. \textbf{B427}, 97 (1998), hep-ph/9712495; C. Giunti,
  hep-ph/9906275; hep-ph/9906456; S.M. Bilenky \textit{et al.}, hep-ph/9907234.

\bibitem{BGKP-96}
S.~M. Bilenky, C.~Giunti, C.~W. Kim, and S.~T. Petcov,
\newblock Phys. Rev. {\bf D54}, 4432 (1996), hep-ph/9604364.

\bibitem{BGG-AB-96}
S.M. Bilenky, C. Giunti and W. Grimus, Proc. of \textit{Neutrino '96},
  Helsinki, June 1996, edited by K. Enqvist \textit{et al.}, p.~174, World
  Scientific, 1997, hep-ph/9609343; Eur. Phys. J. \textbf{C1}, 247 (1998),
  hep-ph/9607372.

\bibitem{Okada-Yasuda-97}
N.~Okada and O.~Yasuda,
\newblock Int. J. Mod. Phys. {\bf A12}, 3669 (1997), hep-ph/9606411.

\bibitem{BGG-bounds-98}
S.~M. Bilenky, C.~Giunti, and W.~Grimus,
\newblock Phys. Rev. {\bf D57}, 1920 (1998), hep-ph/9710209.

\bibitem{BGG-CP-98}
S.~M. Bilenky, C.~Giunti, and W.~Grimus,
\newblock Phys. Rev. {\bf D58}, 033001 (1998), hep-ph/9712537.

\bibitem{BGGS-BBN-98}
S.M. Bilenky, C. Giunti, W. Grimus and T. Schwetz, hep-ph/9804421 [to be
  published in Astropart. Phys.].

\bibitem{BGGS-AB-99}
S.M. Bilenky, C. Giunti, W. Grimus and T. Schwetz, hep-ph/9904316 [to be
  published in Phys. Rev. D].

\bibitem{sterile}
A. Yu. Smirnov, hep-ph/9901208; R.N. Mohapatra, hep-ph/9903261.

\bibitem{Gonzalez-Garcia-atm-analysis-99}
M.~C. Gonzalez-Garcia, H.~Nunokawa, O.~L.~G. Peres, and J.~W.~F. Valle,
\newblock Nucl. Phys. {\bf B543}, 3 (1999), hep-ph/9807305.

\bibitem{Fogli-Lisi-atm}
G.L. Fogli, E. Lisi and D. Montanino, Phys. Rev. \textbf{D49}, 3626 (1994);
  Astrop. Phys. \textbf{4}, 177 (1995); G.L. Fogli, E. Lisi, D. Montanino and
  G. Scioscia, Phys. Rev. \textbf{D55}, 4385 (1997), hep-ph/9607251; G.L.
  Fogli, E. Lisi, A. Marrone and G. Scioscia, Phys. Rev. \textbf{D59}, 033001
  (1999), hep-ph/9808205.

\bibitem{Mikheev-Smirnov-uspekhi-87}
S.~P. Mikheev and A.~Y. Smirnov,
\newblock Sov. Phys. Usp. {\bf 30}, 759 (1987).

\bibitem{Bilenky-Petcov-RMP-87}
S.~M. Bilenky and S.~T. Petcov,
\newblock Rev. Mod. Phys. {\bf 59}, 671 (1987).

\bibitem{Kuo-Pantaleone-RMP-89}
T.~K. Kuo and J.~Pantaleone,
\newblock Rev. Mod. Phys. {\bf 61}, 937 (1989).

\bibitem{CWKim-book}
C.~W. Kim and A.~Pevsner,
\newblock {\em Neutrinos in physics and astrophysics} (Harwood Academic Press,
  Chur, Switzerland, 1993),
\newblock {Contemporary Concepts in Physics, Vol. 8}.

\bibitem{BGG-review-98}
S.~M. Bilenky, C.~Giunti, and W.~Grimus,
\newblock (1998), hep-ph/9812360,
\newblock To be published in Progress in Particle and Nuclear Physics, Volume
  43.

\bibitem{Kuo-Pantaleone-PRL-86}
T.~K. Kuo and J.~Pantaleone,
\newblock Phys. Rev. Lett. {\bf 57}, 1805 (1986).

\bibitem{Shi-Schramm-92}
X.~Shi and D.~N. Schramm,
\newblock Phys. Lett. {\bf B283}, 305 (1992).

\bibitem{Fogli-Lisi-sun}
G.L. Fogli, E. Lisi and D. Montanino, Phys. Rev. \textbf{D49}, 3626 (1994);
  Phys. Rev. \textbf{D54}, 2048 (1996), hep-ph/9605273.

\bibitem{Narayan-three-nu}
Mohan Narayan and G. Rajasekaran and S. Uma Sankar Phys. Rev. \textbf{D53},
  2809 (1996); Phys. Rev. \textbf{D56}, 437 (1997); Phys. Rev. \textbf{D58},
  031301 (1998).

\bibitem{GKM-atm-98}
C.~Giunti, C.~W. Kim, and M.~Monteno,
\newblock Nucl. Phys. {\bf B521}, 3 (1998), hep-ph/9709439.

\bibitem{Akhmedov-Dighe-Lipari-Smirnov-parametric-99}
E.~K. Akhmedov, A.~Dighe, P.~Lipari, and A.~Y. Smirnov,
\newblock Nucl. Phys. {\bf B542}, 3 (1999), hep-ph/9808270.

\bibitem{Barger-variations-98}
V.~Barger, S.~Pakvasa, T.~J. Weiler, and K.~Whisnant,
\newblock Phys. Rev. {\bf D58}, 093016 (1998), hep-ph/9806328.

\bibitem{BBN-Nnu}
S. Burles, K.M. Nollett, J.N. Truran and M.S. Turner, Phys. Rev. Lett.
  \textbf{82}, 4176 (1999), astro-ph/9901157; K.A. Olive, astro-ph/9903309; E.
  Lisi, S. Sarkar and F.L. Villante, Phys. Rev. \textbf{D59}, 123520 (1999),
  hep-ph/9901404; S. Sarkar, astro-ph/9903183.

\bibitem{Foot-Volkas-BBN}
R. Foot, M.J. Thomson and R.R. Volkas, Phys. Rev. \textbf{D53}, 5349 (1996),
  hep-ph/9509327; R. Foot and R.R. Volkas, Phys. Rev. \textbf{D55}, 5147
  (1997), hep-ph/9610229; N.F. Bell, R. Foot and R.R. Volkas, Phys. Rev.
  \textbf{D58}, 105010 (1998), hep-ph/9805259; R. Foot, Astropart. Phys.
  \textbf{10}, 253 (1999), hep-ph/9809315; R. Foot and R.R. Volkas,
  astro-ph/9811067; R. Foot, hep-ph/9906311.

\bibitem{Shi-Fuller-BBN}
X. Shi and G.M. Fuller, Phys. Rev. \textbf{D59}, 063006 (1999),
  astro-ph/9810075; astro-ph/9812232; K. Abazajian, X. Shi and G.M. Fuller,
  astro-ph/{\-}9904052; astro-ph/{\-}9905259.

\bibitem{DiBari-Lipari-Lusignoli-BBN-99}
P. Di Bari, P. Lipari and M. Lusignoli, hep-ph/9907548.

\bibitem{Bugey-95}
Y.~Declais {\em et~al.},
\newblock Nucl. Phys. {\bf B434}, 503 (1995).

\bibitem{Bahcall-WWW}
J. N. Bahcall, {WWW} page: http://{\-}www.{\-}sns.{\-}ias.{\-}edu/{\-}\~{}jnb/.

\bibitem{Baltz-Weneser-earth-87}
A.~J. Baltz and J.~Weneser,
\newblock Phys. Rev. {\bf D35}, 528 (1987).

\bibitem{Baltz-Weneser-earth-94}
A.~J. Baltz and J.~Weneser,
\newblock Phys. Rev. {\bf D50}, 5971 (1994).

\bibitem{Lisi-Montanino-earth-97}
E.~Lisi and D.~Montanino,
\newblock Phys. Rev. {\bf D56}, 1792 (1997), hep-ph/9702343.

\bibitem{Liu-Maris-Petcov-earth1-97}
Q.~Y. Liu, M.~Maris, and S.~T. Petcov,
\newblock Phys. Rev. {\bf D56}, 5991 (1997), hep-ph/9702361.

\bibitem{Petcov-diffractive-98}
S.~T. Petcov,
\newblock Phys. Lett. {\bf B434}, 321 (1998), hep-ph/9805262.

\bibitem{Chizhov-Maris-Petcov-98}
M.~Chizhov, M.~Maris, and S.~T. Petcov,
\newblock (1998), hep-ph/9810501.

\bibitem{Akhmedov-parametric-99}
E.~K. Akhmedov,
\newblock Nucl. Phys. {\bf B538}, 25 (1999), hep-ph/9805272.

\bibitem{Chizhov-Petcov-earth-99}
M.V. Chizhov and S.T. Petcov, hep-ph/9903399; hep-ph/9903424.

\bibitem{Dighe-Liu-Smirnov-earth-99}
A.S. Dighe, Q.Y. Liu and A.Yu. Smirnov, hep-ph/9903329.

\bibitem{Guth-Randall-Serna-earth-99}
A.H. Guth, L. Randall and M. Serna, hep-ph/9903464.

\bibitem{atm-matter}
E.D. Carlson, Phys. Rev. D \textbf{34}, 1454 (1986); P.I. Krastev and S.T.
  Petcov, Phys. Lett. \textbf{B205}, 84 (1988); G. Auriemma, M. Felcini, P.
  Lipari and J.L. Stone, Phys. Rev. \textbf{D37}, 665 (1988); E. Akhmedov, P.
  Lipari and M. Lusignoli, Phys. Lett. \textbf{B300}, 128 (1993).

\bibitem{CDHS-84}
F.~Dydak {\em et~al.},
\newblock Phys. Lett. {\bf 134B}, 281 (1984).

\bibitem{CCFR-84}
I.~E. Stockdale {\em et~al.},
\newblock Phys. Rev. Lett. {\bf 52}, 1384 (1984).

\bibitem{LBL}
M. Campanelli, hep-ex/9905035; P. Picchi and F. Pietropaolo, hep-ph/9812222; K.
  Zuber, hep-ex/9810022; P. Adamson \textit{et al.} (MINOS Coll.), NuMI-L-476
  (March 1999); P. Cennini \textit{et al.} (ICARUS Coll.), LNGS-94/99-I (1994).

\bibitem{nufactory}
S. Geer, Phys. Rev. \textbf{D57}, 6989 (1998), Erratum \textit{ibid.}
  \textbf{D59}, 039903 (1999), hep-ph/9712290; R.N. Mohapatra, hep-ph/9711444;
  B. Autin \textit{et al.}, CERN-SPSC/98-30; A. De Rujula, M.B. Gavela and P.
  Hernandez, Nucl. Phys. \textbf{B547}, 21 (1999), hep-ph/9811390; S. Dutta, R.
  Gandhi and B. Mukhopadhyaya, hep-ph/9905475; V. Barger, S. Geer and K.
  Whisnant, hep-ph/9906487; Slides of the presentations at the ICFA/ECFA
  Workshop on \textit{Neutrino Factories based on Muon Storage Rings},
  $\nu$-Fact'99, Lyon, 5--9 July 1999,
  http://{\-}lyoinfo.{\-}in2p3.{\-}fr/{\-}nufact99/.

\bibitem{DGKK-preparation}
David Dooling, Carlo Giunti, Kyungsik Kang and Chung W. Kim, in preparation.

\end{thebibliography}

\end{document}